\documentclass[onecolumn]{aastex}

\usepackage{amsmath}
\usepackage{mathtools}
\usepackage{tensor}
\usepackage{hyperref}
\hypersetup{colorlinks=true,linkcolor=black,citecolor=black,filecolor=black,urlcolor=black}

\makeatletter
\newcommand{\extref}[1]{\textup{\tagform@{#1}}}
\makeatother

\makeatletter
\def\env@matrix{\hskip -\arraycolsep
  \let\@ifnextchar\new@ifnextchar
  \array{*{\c@MaxMatrixCols}c}}
\makeatother

\newcommand{\dd}{\mathrm{d}}
\newcommand{\pp}{\partial}
\newcommand{\dx}{\dd x}
\newcommand{\dt}{\dd t}
\newcommand{\dr}{\dd r}
\newcommand{\dth}{\dd\theta}
\newcommand{\dph}{\dd\phi}
\newcommand{\ee}{\mathrm{e}}
\newcommand{\abs}[1]{\lvert #1 \rvert}
\newcommand{\altvec}[1]{\mathbf{#1}}

\newcommand{\pgas}{p_\mathrm{gas}}
\newcommand{\pmag}{p_\mathrm{mag}}
\newcommand{\rcrit}{r_\mathrm{c}}
\newcommand{\ucrit}{u_\mathrm{c}}
\newcommand{\Tcrit}{T_\mathrm{c}}
\newcommand{\rin}{r_\mathrm{edge}}
\newcommand{\rpeak}{r_\mathrm{peak}}
\newcommand{\mprot}{m_\mathrm{p}}
\newcommand{\melec}{m_\mathrm{e}}

\shorttitle{GRMHD in Athena++}
\shortauthors{C.~J.~White, J.~M.~Stone, C.~F.~Gammie}

\begin{document}

\title{An Extension of the Athena++ Code Framework for GRMHD Based on Advanced Riemann Solvers and Staggered-Mesh Constrained Transport}
\author{Christopher~J.~White\altaffilmark{1} and James~M.~Stone\altaffilmark{1} and Charles~F.~Gammie\altaffilmark{2,3}}
\altaffiltext{1}{Department of Astrophysical Sciences, Princeton University, 4 Ivy Lane, Princeton, NJ 08544, USA}
\altaffiltext{2}{Department of Physics, University of Illinois, 1110 West Green Street, Urbana, IL 61801, USA}
\altaffiltext{3}{Department of Astronomy, University of Illinois, 1002 West Green Street, Urbana, IL 61801, USA}

\begin{abstract}
  We present a new general relativistic magnetohydrodynamics (GRMHD) code integrated into the Athena++ framework. Improving upon the techniques used in most GRMHD codes, ours allows the use of advanced, less diffusive Riemann solvers, in particular HLLC and HLLD. We also employ a staggered-mesh constrained transport algorithm suited for curvilinear coordinate systems in order to maintain the divergence-free constraint of the magnetic field. Our code is designed to work with arbitrary stationary spacetimes in one, two, or three dimensions, and we demonstrate its reliability in a number of tests. We also report on its promising performance and scalability.
\end{abstract}

\keywords{accretion disks, black hole physics, hydrodynamics, MHD, relativity}

\section{Introduction}
\label{sec:introduction}

The complex, nonlinear dynamics of most astrophysical fluids make numerical simulations a key tool in understanding them, even in flat spacetime. A number of interesting phenomena, however, occur in regions of strong enough gravity that general relativity cannot be neglected, requiring numerical algorithms capable of evolving fluids in curved spacetimes. Examples include accretion onto black holes \citep{Abramowicz2013}, collapsar models of long-duration gamma-ray bursts \citep{Woosley1993}, and merging neutron star binaries \citep{Faber2012}, among others.

Some of the most successful numerical algorithms for solving the equations of compressible fluid dynamics are finite-volume methods, primarily due to their superior accuracy and stability for shock capturing \citep{VanLeer1979}. In such methods the domain is partitioned into discrete cells and the quantities stored and evolved are cell volume averages of the conserved quantities. In Godunov schemes, fluxes of conserved quantities are determined by solving Riemann problems at each interface. In particular, one considers the interface to be a plane separating two spatially constant fluids filling all space, and the solutions of the Riemann problem determine the time-averaged fluxes across the plane in this simplified problem.

A number of Godunov scheme codes have been developed with general relativistic magnetohydrodynamics (GRMHD) capabilities on a stationary background spacetime, including Harm \citep{Gammie2003}, \citet{Komissarov2004}, the Valencia group's code \citep{Anton2006}, and ECHO \citep{DelZanna2007}. In addition, some codes combine the GRMHD and Einstein equations, evolving a self-gravitating magnetized fluid. These include the Tokyo/Kyoto group's code \citep{Shibata2005}, \citet{Anderson2006}, WhiskyMHD \citep{Giacomazzo2007}, GRHydro \citep{Mosta2014}, SpEC \citep{Muhlberger2014}, and IllinoisGRMHD \citep{Etienne2015}.

Two key algorithmic components of a finite-volume Godunov scheme for MHD are the Riemann solver and the method by which the divergence-free constraint on the magnetic field is enforced. The accuracy of the solution depends critically on the accuracy of the Riemann solver adopted. Exact Riemann solvers are however computationally expensive, and the simplest approximate solvers tend to be much more diffusive than their more exact counterparts for subsonic flows. Most of the codes used to date employ simple solvers such as local Lax-Friedrichs or HLLE (see \S\ref{sec:algorithm:riemann} for descriptions); of the aforementioned ten, only \citet{Komissarov2004} and \citet{Anton2006} describe the use of Roe-type Riemann solvers (with the latter making the point that they can in principle use any Riemann solver), and only \citet{Shibata2005} and \citet{Anton2006} describe the use of central, non-Godunov schemes. All others restrict themselves to HLLE and local Lax-Friedrichs. Meanwhile, efficient, approximate solvers that capture some but not all of the wave structure between the outermost waves in the wavefan have been developed for relativistic fluids \citep{Mignone2005,Mignone2009}. In this paper we describe how carefully chosen frame transformations can be used to incorporate such special relativistic algorithms into a general relativistic code, following the suggestion by \citet{Pons1998} and \citet{Anton2006}.

In the case of MHD enforcing the divergence-free constraint is important to prevent spurious production of magnetic monopoles. A number of algorithms have been developed to this end, including the following:
\begin{enumerate}
  \item Divergence cleaning. Auxiliary equations are introduced in order to damp and/or advect to the boundaries any monopoles that are developed. These equations can be elliptic \citep{Brackbill1980,Ramshaw1983}, parabolic \citep{Marder1987}, or hyperbolic \citep{Dedner2002}, with only the last naturally able to respect causality in relativistic settings.
  \item Vector potential evolution. The vector potential components $A^i$ are evolved instead of the magnetic field components $B^i$. This naturally obeys the divergence-free constraint if the vector potential is appropriately staggered, though one must be careful to choose an appropriate gauge, especially when including mesh refinement \citep{Etienne2012}.
  \item Constrained transport (CT). The magnetic field is staggered in such a way as to maintain a specific discretized version of the constraint. \Citet{Evans1988} introduce staggered-mesh CT in a general-relativistic context.
  \item Flux-CT. \Citet{Toth2000} discusses how to smooth the fluxes so as to preseve the divergence-free constraint on a particular stencil of cell-centered magnetic fields.
\end{enumerate}
We choose to employ the \citeauthor{Evans1988}\ method. The details of the algorithm are developed for Cartesian grids by \citet{Gardiner2005}, where only magnetic fields (and not velocities as in the original formulation) are located at interfaces between cells. Here we extend the \citeauthor{Gardiner2005}\ algorithm to arbitrary stationary coordinate systems. Of the aforementioned ten codes, \citet{Anderson2006} uses hyperbolic divergence cleaning, IllinoisGRMHD uses a staggered vector potential, Harm and GRHydro use flux-CT, and the remaining six use a staggered CT scheme.

In this paper we describe a new numerical algorithm for GRMHD that we have developed and implemented in the Athena++ framework, a complete rewrite of the Athena MHD code \citep{Stone2008} that adopts a flexible design necessary for allowing Newtonian, special relativistic, and general relativistic dynamics all within the same code. We will first review the governing equations and their discretizations (\S\ref{sec:equations}) before detailing the steps in our method (\S\ref{sec:algorithm}). We then demonstrate our code's reliability in a number of tests (\S\ref{sec:tests}), as well as full simulations of accretion onto a spinning black hole (\S\ref{sec:torus}). Statistics demonstrating the high-performance nature of the code are given in \S\ref{sec:performance}. In \S\ref{sec:summary} we summarize and conclude our presentation of what is to our knowledge the only 3D finite-volume Godunov GRMHD code that includes both advanced approximate Riemann solvers and staggered-mesh CT, despite the successful use of these techniques in Newtonian and special-relativistic codes such as Athena \citep{Stone2008,Beckwith2011}.

Throughout we use units where $G$ and $c$ are set to unity and a factor of $\sqrt{4\pi}$ has been absorbed into all components of the electromagnetic field. Our metric sign convention, $({-}, {+}, {+}, {+})$, is the same as in \citet{MTW}.

\section{Equations}
\label{sec:equations}

The relevant equations for GRMHD are derived in numerous sources in the literature \citep[e.g.][]{Gammie2003}. Here we list only the most important results, both in order to remove any possible notational ambiguity and to make clear how our particular discretization is implemented.

Consider a perfect fluid with comoving mass density $\rho$, comoving gas pressure $\pgas$, comoving enthalpy per unit mass $h$, and coordinate-frame $4$-velocity components $u^\mu$. For an adiabatic gas of index $\Gamma$, we have $h = 1 + \Gamma/(\Gamma-1) \times \pgas/\rho$. Given magnetic field components $B^i$ in the coordinate frame, define the contravariant field components
\begin{subequations} \begin{align}
  b^0 & = g_{i\mu} B^i u^\mu, \\
  b^i & = \frac{1}{u^0} (B^i + b^0 u^i)
\end{align} \end{subequations}
in terms of the metric and $4$-velocity. The ideal MHD stress-energy tensor then has components
\begin{equation} \label{eq:stress_energy}
  T^{\mu\nu} = (\rho h + b_\lambda b^\lambda) u^\mu u^\nu + \left(\pgas + \frac{1}{2} b_\lambda b^\lambda\right) g^{\mu\nu} - b^\mu b^\nu.
\end{equation}

The governing equations of ideal MHD are
\begin{subequations} \label{eq:evolution} \begin{align}
  \nabla_\mu (\rho u^\mu) & = 0, \label{eq:mass_evolution} \\
  \nabla_\mu T^{\mu\nu} & = 0, \label{eq:momentum_evolution} \\
  \nabla_\mu \tensor[^*]{F}{^{\mu\nu}} & = 0 \label{eq:field_evolution}
\end{align} \end{subequations}
\citep[cf.][1, 3, 14]{Gammie2003}.
Here
\begin{equation} \label{eq:field_tensor}
  \tensor[^*]{F}{^{\mu\nu}} = b^\mu u^\nu - b^\nu u^\mu
\end{equation}
are the components of the dual of the electromagnetic field tensor. Note that there are different sign conventions for the field tensor. In our convention, the components in terms of electric and magnetic fields are
\begin{equation} \label{eq:field_tensor_expanded}
  \tensor[^*]{F}{^{\mu\nu}} =
  \begin{pmatrix}
    0   & -B^1           & -B^2           & -B^3           \\
    B^1 & 0              & \phantom{-}E^3 & -E^2           \\
    B^2 & -E^3           & 0              & \phantom{-}E^1 \\
    B^3 & \phantom{-}E^2 & -E^1           & 0
  \end{pmatrix},
\end{equation}
where $\mu$ indexes rows and $\nu$ columns. As we consider only ideal MHD here, the electric fields can always be inferred from the magnetic fields and fluid velocities via \eqref{eq:field_tensor}.

Our differential equations \eqref{eq:evolution} can be rewritten in terms of partial rather than covariant derivatives, at the expense of introducing some source terms related to the connection coefficients $\Gamma^\sigma_{\mu\nu}$. Splitting timelike and spatial derivatives, we have the flux-conservative equations
\begin{subequations} \label{eq:differential} \begin{align}
  \pp_0 (\sqrt{-g} \rho u^0) + \pp_j (\sqrt{-g} \rho u^j) & = 0, \\
  \pp_0 (\sqrt{-g} \tensor{T}{^0_\mu}) + \pp_j (\sqrt{-g} \tensor{T}{^j_\mu}) & = \sqrt{-g} \Gamma^\sigma_{\rho\mu} \tensor{T}{^\rho_\sigma}, \\
  \pp_0 (\sqrt{-g} B^i) + \pp_j (\sqrt{-g} \tensor[^*]{F}{^{ij}}) & = 0,
\end{align} \end{subequations}
corresponding to \extref{2}, \extref{4}, and \extref{18} of \citet{Gammie2003}.

Consider a cell $V_{i,j,k}$ with $x^1$-, $x^2$-, and $x^3$-indices $i$, $j$, and $k$. Assuming the metric to be stationary, we can assign it the time-independent (but coordinate-dependent) volume
\begin{equation} \label{eq:volume}
  \Delta V_{i,j,k} = \smashoperator{\int\limits_{V_{i,j,k}}} \sqrt{-g} \, \dx^1\,\dx^2\,\dx^3,
\end{equation}
and we can assign its six faces the areas
\begin{subequations} \begin{align}
  \Delta A_{i\pm1/2,j,k} & = \smashoperator{\int\limits_{A_{i\pm1/2,j,k}}} \sqrt{-g} \, \dx^2\,\dx^3, \\
  \Delta A_{i,j\pm1/2,k} & = \smashoperator{\int\limits_{A_{i,j\pm1/2,k}}} \sqrt{-g} \, \dx^1\,\dx^3, \\
  \Delta A_{i,j,k\pm1/2} & = \smashoperator{\int\limits_{A_{i,j,k\pm1/2}}} \sqrt{-g} \, \dx^1\,\dx^2.
\end{align} \end{subequations}
For a conserved quantity $C \in \{\rho u^0, \tensor{T}{^0_\mu}\}$, the volume average stored on the grid at time level $n$ is
\begin{equation}
  C_{i,j,k}^n = \frac{1}{\Delta V_{i,j,k}} \smashoperator[r]{\int\limits_{V_{i,j,k}}} C \sqrt{-g} \, \dx^1\,\dx^2\,\dx^3.
\end{equation}
For the corresponding fluxes $F \in \{\rho u^1, \tensor{T}{^1_\mu}\}$, $G \in \{\rho u^2, \tensor{T}{^2_\mu}\}$, and $H \in \{\rho u^3, \tensor{T}{^3_\mu}\}$, the area averages returned by the Riemann solver are
\begin{subequations} \begin{align}
  F_{i\pm1/2,j,k}^{n+1/2} & = \smashoperator{\int\limits_{A_{i\pm1/2,j,k}}} F \sqrt{-g} \, \dx^2\,\dx^3, \\
  G_{i,j\pm1/2,k}^{n+1/2} & = \smashoperator{\int\limits_{A_{i,j\pm1/2,k}}} G \sqrt{-g} \, \dx^1\,\dx^3, \\
  H_{i,j,k\pm1/2}^{n+1/2} & = \smashoperator{\int\limits_{A_{i,j,k\pm1/2}}} H \sqrt{-g} \, \dx^1\,\dx^2.
\end{align} \end{subequations}
For the corresponding source term $S \in \{0, \Gamma^\sigma_{\rho\mu} \tensor{T}{^\rho_\sigma}\}$, the nonzero volume averages are taken to be
\begin{equation} \label{eq:sources}
  S_{i,j,k}^{n+1/2} = \Gamma^\sigma_{\rho\mu} (\tensor{\bar{T}}{^\rho_\sigma})_{i,j,k}^{n+1/2},
\end{equation}
where the volume-averaged stress-energy components $(\tensor{\bar{T}}{^\rho_\sigma})_{i,j,k}^{n+1/2}$ are found by applying \eqref{eq:stress_energy} to the primitive quantities obtained from inverting the volume-averaged conserved quantities at the appropriate time level (see \S\ref{sec:algorithm:inversion}), with all geometric terms evaluated at the cell center.

With the above definitions, the discretized form of the two hydrodynamical equations in \eqref{eq:differential} can be written
\begin{equation} \begin{split} \label{eq:cons_update}
  C_{i,j,k}^{n+1} & = C_{i,j,k}^n + \Delta t S_{i,j,k}^{n+1/2} \\
  & \quad \qquad + \frac{\Delta t}{\Delta V_{i,j,k}} \left(\Delta A_{i-1/2,j,k} F_{i-1/2,j,k}^{n+1/2} - \Delta A_{i+1/2,j,k} F_{i+1/2,j,k}^{n+1/2}\right) \\
  & \quad \qquad + \frac{\Delta t}{\Delta V_{i,j,k}} \left(\Delta A_{i,j-1/2,k} G_{i,j-1/2,k}^{n+1/2} - \Delta A_{i,j+1/2,k} G_{i,j+1/2,k}^{n+1/2}\right) \\
  & \quad \qquad + \frac{\Delta t}{\Delta V_{i,j,k}} \left(\Delta A_{i,j,k-1/2} H_{i,j,k-1/2}^{n+1/2} - \Delta A_{i,j,k+1/2} H_{i,j,k+1/2}^{n+1/2}\right),
\end{split} \end{equation}
where $\Delta t$ is the change in the timelike coordinate $x^0$ and time level $n+1/2$ is somewhere between $n$ and $n+1$, possibly the same as $n$, depending on the substep of the integration (see \S\ref{sec:algorithm:integrator}).

In the case of magnetic fields, we take the fundamental variables to be area averages. This is a crucial part of the CT scheme of \citet{Evans1988} (see \S\ref{sec:algorithm:ct}). The positioning of quantities and orientations of surfaces and edges is the same as Figure~1 of \citet{Stone2008}. For concreteness, consider the evolution of
\begin{equation}
  (B^1)_{i-1/2,j,k}^n = \frac{1}{\Delta A_{i-1/2,j,k}} \int\limits_{A_1} B^1 \sqrt{-g} \, \dx^2\,\dx^3.
\end{equation}
We define the edge lengths analogously to areas and volumes,
\begin{subequations} \begin{align}
  \Delta L_{i-1/2,j,k\pm1/2} = \smashoperator{\int\limits_{L_{i-1/2,j,k\pm1/2}}} \sqrt{-g} \, \dx^2, \\
  \Delta L_{i-1/2,j\pm1/2,k} = \smashoperator{\int\limits_{L_{i-1/2,j\pm1/2,k}}} \sqrt{-g} \, \dx^3.
\end{align} \end{subequations}
Given the electric fields $E^i$ defined along the edges (as found in \S\ref{sec:algorithm:ct}), we can update the magnetic field according to
\begin{equation} \begin{split} \label{eq:b_update}
  (B^1)_{i-1/2,j,k}^{n+1} & = (B^1)_{i-1/2,j,k}^n \\
  & \quad \qquad + \frac{\Delta t}{\Delta A_{i-1/2,j,k}} \Big(\Delta L_{i-1/2,j-1/2,k} (E^3)_{i-1/2,j-1/2,k}^{n+1/2} \\
  & \quad \qquad \qquad - \Delta L_{i-1/2,j+1/2,k} (E^3)_{i-1/2,j+1/2,k}^{n+1/2}\Big) \\
  & \quad \qquad + \frac{\Delta t}{\Delta A_{i-1/2,j,k}} \Big(\Delta L_{i-1/2,j,k-1/2} (-E^2)_{i-1/2,j,k-1/2}^{n+1/2} \\
  & \quad \qquad \qquad - \Delta L_{i-1/2,j,k+1/2} (-E^2)_{i-1/2,j,k+1/2}^{n+1/2}\Big),
\end{split} \end{equation}
which is the discretized form of the third evolution equation \eqref{eq:differential}.

\section{Numerical Algorithm}
\label{sec:algorithm}

\subsection{Variable Inversion}
\label{sec:algorithm:inversion}

One of the most challenging aspects of numerical methods for relativistic fluid dynamics is variable inversion, the process of extracting the primitive variables from the conserved variables. The difficulty lies in the conserved variables being given by nonlinear functions of the primitive variables with no simple inverses. It is compounded by the fact that only certain primitive states are physically admissible (for example there can be no negative densities or superluminal velocities), and so a given set of conserved variables might not exactly correspond to any allowable set of primitives. Our methods described here are largely based on the $1D_W$ inversion scheme of \citet{Noble2006}.

If magnetic fields are present, they are interpolated to the volume-averaged centers as follows, using the $B^1$ case for concreteness. Given $B^1_{i\pm1/2}$ at interface coordinates $x^1_{i\pm1/2}$ and volume-averaged coordinate $x^1_i$, define $\lambda = (x^1_i - x^1_{i-1/2}) / (x^1_{i+1/2} - x^1_{i-1/2})$. Then the volume-averaged value of $B^1$ is taken to be
\begin{equation} \label{eq:b_interpolation}
  B^1_i = (1 - \lambda) B^1_{i-1/2} + \lambda B^1_{i+1/2}.
\end{equation}

In the case of general relativity, we take the primitive variables to be $\{\rho, \pgas, \tilde{u}^i\}$ and the conserved variables to be $\{\rho u^0, \tensor{T}{^0_\mu}\}$. Here we use the same velocities as in \citet{Noble2006}: $\tilde{u}^\mu = (\delta^\mu_\nu + n^\mu n_\nu) u^\nu$, where $n_\mu = -\alpha \delta^0_\mu$ are the components of the future-pointing unit vector normal to constant-$x^0$ hypersurfaces, and $\alpha = (-g^{00})^{-1/2}$ is the lapse. These projections have the desirable property that they describe subluminal motion no matter what values they have. Additionally, the fourth component is $\tilde{u}^0 = 0$. The Lorentz factor of the fluid as seen by the normal observer is $\gamma = (1 + g_{ij} \tilde{u}^i \tilde{u}^j)^{1/2}$. Following the $1D_W$ scheme, the Newton-Raphson method is used to solve a nonlinear equation for the relativistic gas enthalpy $W = \gamma^2 \rho h$. In the hydrodynamics case we use the same procedure as with MHD, simply setting all magnetic fields to $0$.

In special relativity, we take the primitive variables to be $\{\rho, \pgas, v^i\}$ and the conserved variables to be $\{D, E, M^i\} = \{\gamma \rho, T^{0\mu}\}$, where $\gamma = u^0$ is the fluid Lorentz factor in the coordinate frame. For MHD, we also use the Newton-Raphson method to find $W$, where we simply have $W = \gamma^2 \rho h$. The method is detailed in \citet{Mignone2007}, though we iterate on the residual of $E$ as a function of $W$ rather than $E - D$ as a function of $W - D$ as they do.

In the particular case of special relativistic hydrodynamics, we can avoid using an iterative method and instead directly solve a quartic equation \citep{Schneider1993}. In particular, the $3$-velocity magnitude $\abs{v}$ must satisfy
\begin{equation}
  a_4 \abs{v}^4 + a_3 \abs{v}^3 + a_2 \abs{v}^2 + a_1 \abs{v} + a_0 = 0,
\end{equation}
where in terms of $M^2 = \eta_{ij} M^i M^j$ and $\abs{M} = \sqrt{M^2}$ we define
\begin{subequations} \begin{align}
  a_4 & = (\Gamma-1)^2 (D^2 - M^2), \\
  a_3 & = -2 \Gamma (\Gamma-1) \abs{M} E, \\
  a_2 & = \Gamma^2 E^2 + 2 (\Gamma-1) M^2 - (\Gamma-1)^2 D^2, \\
  a_1 & = -2 \Gamma \abs{M} E, \\
  a_0 & = M^2.
\end{align} \end{subequations}
Given the root $\abs{v}$, we recover the primitives
\begin{subequations} \begin{align}
  \rho & = D \sqrt{1-\abs{v}^2}, \\
  v^i & = \frac{\abs{v}}{\abs{M}} M^i, \\
  \pgas & = (\Gamma-1) (E - \eta_{ij} M^i v^j - \rho).
\end{align} \end{subequations}

As part of variable inversion we impose limits on the recovered variables. Since negative densities and pressures are unphysical, and since vanishing values cause problems in finite-volume fluid codes, we ensure $\rho > \rho_\mathrm{min}$ and $\pgas > \pgas^\mathrm{min}$, where $\rho_\mathrm{min}$ and $\pgas^\mathrm{min}$ are adjusted based on the problem. We also must obey the physical constraint $\gamma \geq 1$, and for numerical purposes we limit $\gamma < \gamma_\mathrm{max}$, where $\gamma_\mathrm{max}$ is often chosen to be $100$.

If the recovered primitives fall outside these ranges, they are modified to the nearest admissible values. The primitives are also set to the floor values in cases where the iteration does not converge. Any such modifications are then followed by reevaluating the conserved quantities in the given cell.

\subsection{Reconstruction}
\label{sec:algorithm:reconstruction}

As input to the Riemann problem, one needs fluid states on either side of the interface. While one can simply take the states from the two adjacent cells, this method is only first-order accurate in the size of the cells. For smooth flows, information from a larger stencil can be used to infer left and right states in a higher-order way. Reconstruction is the process of interpolating these states while preserving any actual discontinuities and not introducing spurious extrema.

For a primitive quantity $q$, we define the values on the left and right sides of the $x^1_{i-1/2}$ interface using piecewise linear reconstruction. Specifically, we use the modified van~Leer slope limiter algorithm of \citet{Mignone2014}, which reconstructs $q_{i-1/2,\mathrm{L/R}}$ from the values $q_{i-2}$, $q_{i-1}$, $q_i$, and $q_{i+1}$. This algorithm correctly takes into account nonuniform spacing, as well as the difference between the volume-averaged cell center coordinates (``centroids of volume'' in that work) and arithmetic means of surface coordinates (``geometrical cell centers''). The slope limiter has been proved to be total variation diminishing (TVD) for all orthogonal coordinate systems. While this guarantee has not been extended to non-orthogonal coordinate systems, to our knowledge this has not been done with any other method, and indeed a number of methods in use are not even TVD with orthogonal but non-Cartesian or non-uniform grids.

For completeness, we summarize the \citeauthor{Mignone2014}\ algorithm here. Define
\begin{subequations} \begin{align}
  \Delta_\mathrm{L} & = \frac{q_{i-1}-q_{i-2}}{x^1_{i-1}-x^1_{i-2}}, \\
  \Delta_\mathrm{C} & = \frac{q_i-q_{i-1}}{x^1_i-x^1_{i-1}}, \\
  \Delta_\mathrm{R} & = \frac{q_{i+1}-q_i}{x^1_{i+1}-x^1_i}.
\end{align} \end{subequations}
Define also
\begin{subequations} \begin{align}
  \Delta x_\mathrm{L} & = x^1_{i-1/2} - x^1_{i-1}, & \Delta x_\mathrm{R} & = x^1_i - x^1_{i-1/2}, \\
  C^\mathrm{F}_\mathrm{L} & = \frac{x^1_i-x^1_{i-1}}{\Delta x_\mathrm{L}}, & C^\mathrm{F}_\mathrm{R} & = \frac{x^1_{i+1}-x^1_i}{x^1_{i+1/2}-x^1_i}, \\
  C^\mathrm{B}_\mathrm{L} & = \frac{x^1_{i-1}-x^1_{i-2}}{x^1_{i-1}-x^1_{i-3/2}}, & C^\mathrm{B}_\mathrm{R} & = \frac{x^1_i-x^1_{i-1}}{\Delta x_\mathrm{R}}.
\end{align} \end{subequations}
The left and right quantities are then given by
\begin{subequations} \begin{align}
  q_{i-1/2,\mathrm{L}} & =
  \begin{cases}
    \displaystyle q_{i-1}, & \Delta_\mathrm{L} \Delta_\mathrm{C} \leq 0, \\
    \displaystyle q_{i-1} + \frac{\Delta x_\mathrm{L}\Delta_\mathrm{L}\Delta_\mathrm{C}(C^\mathrm{F}_\mathrm{L}\Delta_\mathrm{L}+C^\mathrm{B}_\mathrm{L}\Delta_\mathrm{C})}{\Delta_\mathrm{L}^2+(C^\mathrm{F}_\mathrm{L}+C^\mathrm{B}_\mathrm{L}-2)\Delta_\mathrm{L}\Delta_\mathrm{C}+\Delta_\mathrm{C}^2}, & \Delta_\mathrm{L} \Delta_\mathrm{C} > 0;
  \end{cases} \\
  q_{i-1/2,\mathrm{R}} & =
  \begin{cases}
    \displaystyle q_i, & \Delta_\mathrm{C} \Delta_\mathrm{R} \leq 0, \\
    \displaystyle q_i - \frac{\Delta x_\mathrm{R}\Delta_\mathrm{C}\Delta_\mathrm{R}(C^\mathrm{F}_\mathrm{R}\Delta_\mathrm{C}+C^\mathrm{B}_\mathrm{R}\Delta_\mathrm{R})}{\Delta_\mathrm{C}^2+(C^\mathrm{F}_\mathrm{L}+C^\mathrm{B}_\mathrm{L}-2)\Delta_\mathrm{C}\Delta_\mathrm{R}+\Delta_\mathrm{R}^2}, & \Delta_\mathrm{C} \Delta_\mathrm{R} > 0.
  \end{cases}
\end{align} \end{subequations}

In the case of transverse magnetic field $q \in \{B^2, B^3\}$, the cell-centered values are used in the reconstruction algorithm. The longitudinal field $B^1$ is already defined at $x^1_{i-1/2}$, and so no reconstruction is needed for this quantity. Reconstruction for $x^2$- and $x^3$-surfaces proceeds analogously.

\subsection{Frame Transformation}
\label{sec:algorithm:transformation}

Rather than try to solve a Riemann problem directly in general coordinates, we choose to explicitly transform the left and right states at an interface to a locally Minkowski frame. In this way, we can leverage the advances made in special relativistic Riemann problems, as suggested by \citet{Pons1998} and \citet{Anton2006}. While the scalar quantities $\rho$ and $\pgas$ transform between frames trivially, vector quantities rely on constructing a new basis $\{\vec{e}_{(\hat{\mu})}\}$ from the old coordinate basis $\{\vec{\pp}_{(\mu)}\}$, where the inner product with respect to the metric yields $\vec{\pp}_{(\mu)} \cdot \vec{\pp}_{(\nu)} = g_{\mu\nu}$ but $\vec{e}_{(\hat{\mu})} \cdot \vec{e}_{(\hat{\nu})} = \eta_{\hat{\mu}\hat{\nu}}$. Note that here we use parentheses around subscripts to indicate indices into generic sets, as opposed to components of tensors.

For a surface of constant $x^i$, we define the new basis to have the following properties:
\begin{enumerate}
  \item $\vec{e}_{(\hat{\mu})}$ must be orthogonal to $\vec{e}_{(\hat{\nu})}$ for all $\mu \neq \nu$.
  \item Each $\vec{e}_{(\hat{\mu})}$ must be normalized to have an inner product of $\pm1$ with itself, with $\vec{e}_{(\hat{t})}$ being timelike and $\vec{e}_{(\hat{\jmath})}$ being spacelike.
  \item $\vec{e}_{(\hat{t})}$ must be orthogonal to surfaces of constant $x^0$.
  \item The projection of $\vec{e}_{(\hat{x})}$ onto the surface of constant $x^0$ must be orthogonal to the surface of constant $x^i$ within that submanifold.
\end{enumerate}
The first two properties are a restatement of the condition $\vec{e}_{(\hat{\mu})} \cdot \vec{e}_{(\hat{\nu})} = \eta_{\hat{\mu}\hat{\nu}}$. Without the third property, the fluxes returned by the Riemann solver would not necessarily correspond to time evolution in the understood $x^0$-direction. The fourth property allows us to ignore $y$- and $z$-fluxes in the Riemann problem, using only the $x$-fluxes returned by a $1$-dimensional Riemann solver to infer the $x^1$-fluxes.

The details of the frame transformation are worked out in the appendix (\S\ref{sec:transformation}). Here we quote the results. For an $x^1$-interface, define the components
\begin{equation} \label{eq:to_global}
  \tensor{M}{^\mu_{\hat{\nu}}} =
  \begin{pmatrix}
    A g^{00} & 0                               & 0                   & 0 \\
    A g^{01} & B (g^{01}g^{01} - g^{00}g^{11}) & 0                   & 0 \\
    A g^{02} & B (g^{01}g^{02} - g^{00}g^{12}) & \phantom{-}D g_{33} & 0 \\
    A g^{03} & B (g^{01}g^{03} - g^{00}g^{13}) & -D g_{23}           & C
  \end{pmatrix},
\end{equation}
where we have the shorthands
\begin{subequations} \begin{align}
  A & = -(-g^{00})^{-1/2}, \\
  B & = \big(g^{00} (g^{00}g^{11} - g^{01}g^{01})\big)^{-1/2}, \\
  C & = (g_{33})^{-1/2}, \\
  D & = \big(g_{33} (g_{22}g_{33} - g_{23}g_{23})\big)^{-1/2}.
\end{align} \end{subequations}
This matrix transforms contravariant tensors in the locally Minkowski basis to the coordinate basis, so for example $\tensor{T}{^1_\mu} = g_{\mu\nu} \tensor{M}{^1_{\hat{\mu}}} \tensor{M}{^\nu_{\hat{\nu}}} T^{\hat{\mu}\hat{\nu}}$ are the $x^1$-fluxes of $4$-momentum in the coordinate basis if $T^{\hat{\mu}\hat{\nu}}$ are the stress-energy tensor components in the orthonormal basis. The inverse transformation takes the form
\begin{equation} \label{eq:to_local}
  \tensor{M}{^{\hat{\mu}}_\nu} =
  \begin{pmatrix}
    -A                                     & 0                                      & 0                     & 0   \\
    B g^{01}                               & -B g^{00}                              & 0                     & 0   \\
    B^2E g^{00} / D g_{33}                 & B^2F g^{00} / D g_{33}                 & 1 / D g_{33}          & 0   \\
    (B^2/C) g^{00} (G + E g_{23} / g_{33}) & (B^2/C) g^{00} (H + F g_{23} / g_{33}) & (1/C) g_{23} / g_{33} & 1/C
  \end{pmatrix},
\end{equation}
where we define
\begin{subequations} \begin{align}
  E & = g^{01} g^{12} - g^{11} g^{02}, \\
  F & = g^{01} g^{02} - g^{00} g^{12}, \\
  G & = g^{01} g^{13} - g^{11} g^{03}, \\
  H & = g^{01} g^{03} - g^{00} g^{13}.
\end{align} \end{subequations}
With this matrix we can for example write the orthonormal frame fluid $4$-velocity components in terms of the coordinate basis components: $u^{\hat{\mu}} = \tensor{M}{^{\hat{\mu}}_\nu} u^\nu$.

These formulas hold for the interfaces of constant $x^2$ (respectively $x^3$) under one (respectively two) applications of the cyclic permutation $1 \mapsto 2 \mapsto 3$ in all indices, including the rows of $\tensor{M}{^\mu_{\hat{\nu}}}$ and the columns of $\tensor{M}{^{\hat{\mu}}_\nu}$.

One important detail to be noted is that the constant-coordinate interface across which we desire fluxes will generally appear to be moving in the orthonormal coordinates. To see this, consider an interface of constant $x^1$ and how it appears in orthonormal coordinates $(\hat{t},\hat{x},\hat{y},\hat{z})$. According to \eqref{eq:to_global} we know
\begin{equation} \label{eq:interface_coordinates}
  x^1 = A g^{01} \hat{t} + B (g^{01} g^{01} - g^{00} g^{11}) \hat{x}.
\end{equation}
Given that $x^1$ is constant, differentiation of \eqref{eq:interface_coordinates} tells us the interface velocity can be calculated as
\begin{equation} \label{eq:interface_velocity}
  v^{\hat{x}} \equiv \frac{\dd\hat{x}}{\dd\hat{t}} = -\frac{Ag^{01}}{B(g^{01}g^{01}-g^{00}g^{11})} = \frac{g^{01}}{\sqrt{g^{01}g^{01}-g^{00}g^{11}}}.
\end{equation}
In the language of the $3{+}1$ formalism with lapse $\alpha$, shift components $\beta^i$, and contravariant $3$-metric components $\gamma^{ij}$, we can write $v^{\hat{x}} = \beta^1 / \alpha \sqrt{\gamma^{11}}$, in agreement with \citet{Pons1998}. That is, a nonzero shift in the direction of interest results in a nonzero interface velocity in the chosen orthonormal frame.

As this transformation yields physical quantities as measured by a normal observer, it can be applied even inside the ergosphere or, in the case of horizon-penetrating coordinates, inside the event horizon. This would not be the case if for example we attempted to transform into the frame of a stationary observer, since the spacetime might not everywhere admit timelike stationary worldlines. The transformation can be singular at coordinate singularities such as a polar axis, but such cases must be handled carefully no matter what method is being employed.

\subsection{Riemann solver}
\label{sec:algorithm:riemann}

The Riemann problem is that of determining time-averaged fluxes across an interface given constant left and right states. A wide variety of exact and approximate solvers have been developed. For example, Roe solvers \citep{Roe1981} solve a full linearized set of equations, obtaining high accuracy at the cost of performance.

An alternative is the HLL family of Riemann solvers, based on the work of \citet{Harten1983}, which have the advantage of being computationally simple while guaranteeing positivity (i.e.\ returning physically admissible intermediate states given physically admissible inputs). All HLL Riemann solvers take as input left and right states as well as leftgoing and rightgoing signal speeds, which we calculate as either the sound speeds in hydrodynamics or the fast magnetosonic speeds in MHD.

\begin{figure}
  \centering
  \includegraphics[width=6in]{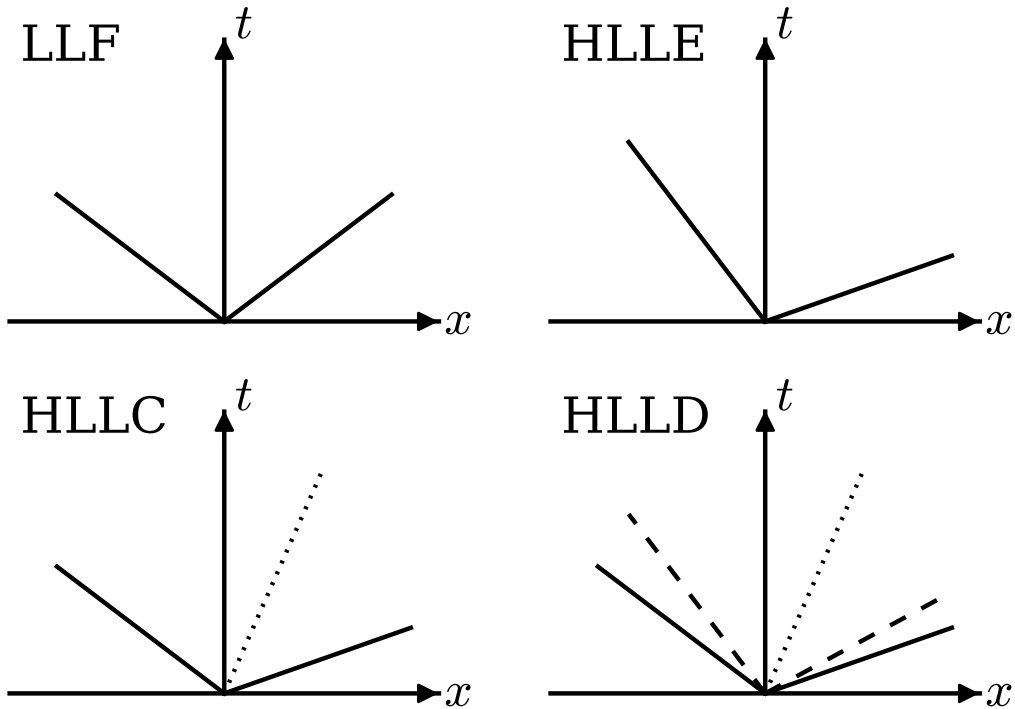}
  \caption{Schematic $1{+}1$ spacetime diagrams of different HLL Riemann solvers' wavefans. The solid diagonal lines indicate the fastest moving waves in either direction, with LLF assuming the speeds are the same. The dotted lines in HLLC and HLLD represent the contact discontinuity, and the dashed lines in HLLD represent the Alfv\'en waves. The ordering of the waves shown is accurate, though waves can be degenerate. Fluxes are determined along the $x = 0$ surface for interfaces at rest, which can be in any of the $3$, $4$, or $6$ regions of the wavefan. Slopes are qualitative; in particular all wavespeeds are subluminal. \label{fig:riemann}}
\end{figure}

A number of different HLL Riemann solvers have been developed for relativistic fluids. We have the following available in the code:
\begin{enumerate}
  \item LLF (local Lax-Friedrichs). The largest in magnitude eigenvalue of the Roe matrix \citep{Roe1981}, i.e.\ the fastest linear wavespeed, is taken to be the signal speed on both sides of the interface. This solver is computationally inexpensive even in relativity, as it has only a single intermediate state and neglects the physical nature of the left and right states. This comes at the cost of being diffusive for flows slower than the extremal wavespeed (see \S\ref{sec:tests:linear}).
  \item HLLE (Harten-Lax-van~Leer+Einfeldt). This solver is nearly identical to LLF, having just a single intermediate state, with the only difference being that the two signal speeds are allowed to be different. We use the speed of the fastest leftgoing wave, considering both sides of the interface, as well as the fastest rightgoing wave, as suggested by \citet{Einfeldt1988}. In practice HLLE and LLF give very similar results and have only small performance differences.
  \item HLLC (HLL+contact). By resolving not only the extremal waves but also the contact discontinuity in the Riemann fan, one can get better results. Such an HLLC Riemann solver is developed for special relativity in \citet{Mignone2005}, and we implement that algorithm for hydrodynamics. \Citet{Mignone2006} also develop an HLLC solver for relativistic MHD, though we forego implementing this in favor of HLLD.
  \item HLLD (HLL+discontinuities). This solver is developed for relativistic MHD by \citet{Mignone2009}. Given the extremal (fast magnetosonic) wavespeeds, it resolves not only the contact but also the Alfv\'en waves. It improves the accuracy of the fluxes by lessening the numerical diffusion, though at the expense of requiring nonlinear root finds. In particular, we use the secant method to find the total pressure across the contact. In the case of a strong longitudinal magnetic field, the conserved HLLE state is constructed and inverted using a Newton-Raphson solver in order to find a total pressure to initialize the secant method.
\end{enumerate}
Figure~\ref{fig:riemann} illustrates how the different solvers treat the internal shock structure.

Owing to the frame transformation, the same Riemann solvers can be used in both special and general relativity. As noted in \S\ref{sec:algorithm:transformation}, however, one must be careful to consider the Riemann problem with a moving interface. This is done by constructing the waves and intermediate states as usual and then determining the relevant portion of the wavefan according to where the generally nonzero interface velocity \eqref{eq:interface_velocity} falls relative to the wavespeeds. In the schematic sense of Figure~\ref{fig:riemann}, the fluxes sought are not necessarily along the vertical lines, but could be along angled lines also passing through the origin.

Additionally, the transformation \eqref{eq:to_global} back to global coordinates will have a nonzero component $\tensor{M}{^1_{\hat{t}}}$, in addition to the ever-present $\tensor{M}{^1_{\hat{x}}}$, whenever there is a moving interface. This necessitates having not just the components $T^{\hat{x}\hat{\nu}}$ but also $T^{\hat{t}\hat{\nu}}$ in the orthonormal frame. That is, we need the conserved state in the appropriate region of the wavefan, not just the associated fluxes. In practice even the more complicated Riemann solvers solve for the intermediate states by finding primitive variables, from which one can deduce consistent conserved states as well as fluxes.

The general procedure for obtaining fluxes proceeds as follows. First, the Riemann problem consisting of left and right states, extremal wavespeeds, and an interface velocity is transformed into the appropriate locally Minkowski frame. Next, the Riemann solver is used to solve for the wavefan, including the speeds of internal waves as well as the conserved quantities and their associated fluxes normal to the interface in each region separated by the waves. The appropriate region is then selected based on where the interface velocity falls inside the wavefan. Finally, the fluxes and conserved quantities in that region are transformed back to fluxes in the original coordinate system.

\subsection{Constrained Transport (CT)}
\label{sec:algorithm:ct}

We employ a CT update of the face-centered magnetic fields in order to maintain the divergence-free constraint without resorting to divergence cleaning. The fundamentals of CT were in fact developed within the context of general relativity \citep{Evans1988}. Our implementation differs slightly from that original description, for example by having cell-centered rather than face-centered velocities. Ours is a simple extension of the algorithm detailed in \citet{Gardiner2005} to relativity, and we summarize it here.

\begin{figure}
  \centering
  \includegraphics[width=6in]{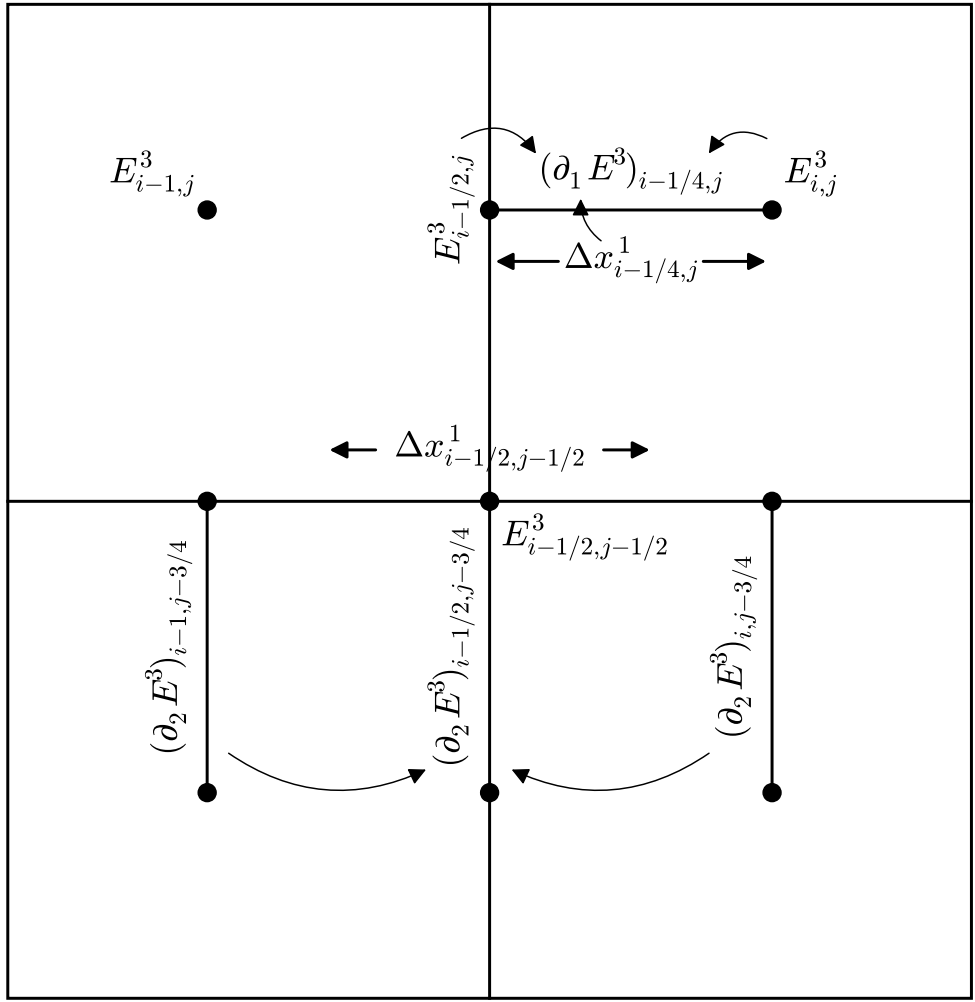}
  \caption{$2{+}0$ spacetime diagram showing a timeslice of four cells. The $x^3$ direction is out of the plane of the figure, and indices $k$ have been suppressed from all quantities. The top left cell shows what is known at the end of the Riemann problem:\ electric fields on the faces, as well as easily-deducible cell-centered electric fields. The top right cell illustrates how a gradient is determined. The bottom two cells show gradients being upwinded back to the faces. Four face-centered fields, four upwinded gradients, and two edge-centered widths are combined to yield the edge-centered electric field $E^3_{i-1/2,j-1/2}$. \label{fig:ct_diagram}}
\end{figure}

The goal of the algorithm is to determine edge-centered electric fields to use in the update \eqref{eq:b_update} consistent with the fluxes returned by the Riemann solver. For concreteness, we will show how we calculate $E^3_{i-1/2,j-1/2}$, dropping all time superscripts. The algorithm proceeds in five steps:
\begin{enumerate}
  \item Obtain the four face-centered electric fields $E^3_{i-1/2,j(-1)}$ and $E^3_{i(-1),j-1/2}$ from the Riemann solver.
  \item Calculate the four cell-centered electric fields $E_{i(-1),j(-1)}$.
  \item Calculate the eight electric field gradients $(\pp_1 E^3)_{i-1/4(-1/2),j(-1)}$ and $(\pp_2 E^3)_{i(-1),j-1/4(-1/2)}$.
  \item Upwind the electric field gradients to obtain the four gradients $(\pp_1 E^3)_{i-1/4(-1/2),j-1/2}$ and $(\pp_2 E^3)_{i-1/2,j-1/4(-1/2)}$.
  \item Combine the face-centered fields and upwinded gradients to get the edge-centered field $E^3_{i-1/2,j-1/2}$.
\end{enumerate}
The relevant quantities are laid out schematically in Figure~\ref{fig:ct_diagram}.

For the first step, we simply note that the electric fields in the $3$-direction are nothing more than the $1,2$-components of the dual of the electromagnetic field tensor \eqref{eq:field_tensor_expanded} and are thus the fluxes returned by the Riemann solver in the $x^1$ and $x^2$ directions (after transforming back to global coordinates). In particular, $E^3_{i-1/2,j(-1)}$ are the negatives of the $x^1$-fluxes of $B^2$, $E^3_{i-1/2,j(-1)} = -\tensor*[^*]{F}{^{21}_{i-1/2,j(-1)}}$. Similarly, we use the positive $x^2$-fluxes of $B^1$ to obtain $E^3_{i(-1),j-1/2} = \tensor*[^*]{F}{^{12}_{i(-1),j-1/2}}$.

For the second step (which does not depend on the first), we calculate the cell-centered electric fields from the cell-centered velocities and the interpolated magnetic fields (e.g.\ \eqref{eq:b_interpolation}). This amounts to simply constructing the appropriate $4$-vector components $u^\mu$ and $b^\mu$ and using \eqref{eq:field_tensor} and \eqref{eq:field_tensor_expanded}.

In the third step we take gradients of electric field components from cell centers to faces, similar to \extref{45} of \citeauthor{Gardiner2005}. In order to calculate such gradients, we require cell widths, though as we shall see these can be cancelled from the final expression to a good approximation. Consider the gradient $\pp_1 E^3$ located between cell center $i,j$ and face center $i-1/2,j$, as in the top right cell of Figure~\ref{fig:ct_diagram}. If we have
\begin{equation}
  \Delta x^1_{i-1/4,j} = \int\limits_{C} \sqrt{g_{11}} \, \dx^1
\end{equation}
with $C$ the curve of constant $x^0$, $x^2$, and $x^3$ running from the cell center to the face, then we can take
\begin{equation} \label{eq:e_grad}
  (\pp_1 E^3)_{i-1/4,j} = \frac{E^3_{i,j}-E^3_{i-1/2,j}}{\Delta x^1_{i-1/4,j}}.
\end{equation}
Similarly we can calculate the other seven such gradients.

Fourth, we shift the gradients onto the faces via upwinding. For example, as illustrated in Figure~\ref{fig:ct_diagram}, we calculate
\begin{equation}
  (\pp_2 E^3)_{i-1/2,j-3/4} =
  \begin{cases}
    (\pp_2 E^3)_{i-1,j-3/4}, & u^1_{i-1/2,j-1} > 0, \\
    (\pp_2 E^3)_{i,j-3/4}, & u^1_{i-1/2,j-1} < 0,
  \end{cases}
\end{equation}
according to the sign of the mass flux returned by the Riemann solver. This is \extref{50} of \citeauthor{Gardiner2005}.

Finally, we compute the edge-centered value as is done in \extref{41} of \citeauthor{Gardiner2005}. If we have an appropriate width $\Delta x^1_{i-1/2,j-1/2}$ defined similarly to $\Delta x^1_{i-1/4,j}$---that is, by integrating $\sqrt{g_{11}}$ as $x^1$ varies over half a cell, as shown in Figure~\ref{fig:ct_diagram}---then the value we seek is
\begin{equation} \begin{split} \label{eq:e_combined}
  E^3_{i-1/2,j-1/2} & = \frac{1}{4} \left(E^3_{i-1/2,j} + E^3_{i-1/2,j-1} + E^3_{i,j-1/2} + E^3_{i-1,j-1/2}\right) \\
  & \quad \qquad + \frac{\Delta x^1_{i-1/2,j-1/2}}{4} \left(E^3_{i-3/4,j-1/2} - E^3_{i-1/4,j-1/2}\right) \\
  & \quad \qquad + \frac{\Delta x^2_{i-1/2,j-1/2}}{4} \left(E^3_{i-1/2,j-3/4} - E^3_{i-1/2,j-1/4}\right).
\end{split} \end{equation}
However, if the metric and grid spacing are varying smoothly, we can approximately cancel the $\Delta x$ factors in \eqref{eq:e_combined} with those in \eqref{eq:e_grad}.

\subsection{Source Terms}
\label{sec:algorithm:sources}

In non-Cartesian coordinates one must consider not only fluxes but also geometric source terms when updating conserved quantities according to \eqref{eq:cons_update}. As can be seen in \eqref{eq:differential}, the continuity and magnetic field equations never have source terms but the energy-momentum equation generally does.

As noted in \citet{Gammie2003}, with the adopted index placement on the stress-energy tensor the source terms for $\tensor{T}{^0_\mu}$ will vanish when the metric does not depend on $x^\mu$. In particular, since the metric is stationary we never have geometric source terms in energy.

For the remaining momentum equations we evaluate the stress-energy components using the primitive variables at the appropriate timestep, taking the metric to have its cell-centered values. We then contract these components with the connection coefficients according to \eqref{eq:sources}, resulting in the appropriate quantities for adding to the conserved variables.

Source terms for additional physics (e.g.\ heating and cooling functions) can be added to the conserved quantities along with the geometric terms.

\subsection{Time Integrator}
\label{sec:algorithm:integrator}

At the heart of the numerical implementation is the time integrator. We employ a temporally second-order van~Leer integrator for all quantities. Given the conserved variables at time step $n$, we first calculate the associated primitive variables (\S\ref{sec:algorithm:inversion}). Next, Riemann problems are set up (\S\ref{sec:algorithm:reconstruction} and \S\ref{sec:algorithm:transformation}) and solved (\S\ref{sec:algorithm:riemann}), yielding fluxes. We also compute any necessary source terms from the primitives (\S\ref{sec:algorithm:sources}).

The fluxes and source terms are used to update all conserved variables by half a timestep, including the magnetic fields (\S\ref{sec:algorithm:ct}). The same procedure is repeated at time step $n{+}1/2$, except the resulting fluxes and source terms are used to update the step $n$ state to step $n+1$. Schematically, we advance the grid by a single timestep via
\begin{subequations} \begin{align}
  C^{n+1/2} & = C^n + \frac{\Delta t}{2} S^n + \frac{\Delta t}{2\Delta V} \sum_\mathrm{faces} F^n \Delta A, \\
  C^{n+1} & = C^n + \Delta t S^{n+1/2} + \frac{\Delta t}{\Delta V} \sum_\mathrm{faces} F^{n+1/2} \Delta A.
\end{align} \end{subequations}
The van~Leer integrator is TVD, and so it will not introduce spurious extrema.

Athena++ implements a variety of other temporal integration algorithms, including the TVD second-order and third-order Runge-Kutta methods of \citet{Shu1988}. The RK2 integrator is the same as Heun's method and we have found is no better than the van~Leer method we implement. It suffers from the drawback that second-order spatial reconstruction in both substeps, whereas we can use first-order reconstruction for the first half step with the van~Leer scheme and still obtain second-order convergence. RK3 and other higher-order schemes generally require extra storage of intermediate results, and for our present purposes we find second order in time to be sufficient. CT corner transport upwind (CTU) integrators have also been developed for MHD \citep{Gardiner2005,Gardiner2008}, but CTU methods require time-advanced variable estimates that are difficult to generalize to arbitrary coordinate systems.

\subsection{Storage Requirements}
\label{sec:algorithm:storage}

At each cell, we must store $5$ primitive variables and $5$ conserved variables for each half step as required by the van~Leer integrator. For MHD in $3$ spatial dimensions we also store $3$ each of face-centered and cell-centered magnetic fields and edge-centered, face-centered, and cell-centered electric fields, all at each half step. This amounts to $20$ numbers per cell in hydrodynamics, and $50$ numbers per cell in MHD.

Any finite volume code must have access to cell volumes, interface areas, edge lengths (in the case of MHD), and cell widths (to determine maximum stable timesteps). In general relativity, the values of $g_{\mu\nu}$, $g^{\mu\nu}$, $\Gamma^\sigma_{\mu\nu}$, and the transformation matrices \eqref{eq:to_global} and \eqref{eq:to_local} are also required at various stages of the integration. Fortunately the assumption of a stationary metric means these terms can be precomputed. Moreover, symmetries of the metric combined with the restriction that cells be divided by interfaces of constant coordinate values often results in separable formulas for these values, obviating the need for 3D storage arrays.

For example, in the Kerr-Schild coordinates used in \S\ref{sec:tests:inflow} and \S\ref{sec:torus}, the volume \eqref{eq:volume} of a cell bounded by coordinates $r_\pm$, $\theta_\pm$, and $\phi_\pm$ is
\begin{equation}
  \Delta V = \frac{1}{3} (r_+-r_-) (\cos\theta_--\cos\theta_+) (\phi_+-\phi_-) \left(f(r_-,r_+) + a^2 f(\cos\theta_-,\cos\theta_+)\right),
\end{equation}
where $f(x,y) = x^2 + xy + y^2$. The three parenthesized prefactors and the two $f$ terms can each be stored in 1D arrays, where $\Delta V$ is computed using just a small handful of additions and multiplications when needed. In fact, no more than 2D storage is required for any metric discussed in this work. For the most complicated metric discussed here, Kerr-Schild, we store $22 N_r + 32 N_\theta + 9 N_\phi + 20 N_r N_\theta$ values on an $N_r \times N_\theta \times N_\phi$ grid. This may seem substantial in comparison to the $4 N_x + 4 N_y + 4 N_z$ coordinate-related values stored when using the Minkowski metric, but still pales in comparison to the $(20\text{--}50) N_1 N_2 N_3$ values holding evolving variables. Thus the memory footprint of our code is substantially smaller than that of codes that store all geometric factors without regard for symmetry or separability.

\section{Tests}
\label{sec:tests}

Because much of the machinery for our code is the same in both general and special relativity, several of our tests are special relativistic in nature. Some are run in both special and general relativistic settings, and others are purely general relativistic.

\subsection{Linear Waves}
\label{sec:tests:linear}

Linear wave convergence is a strong test of a code, and so we present convergence results for special relativistic hydrodynamics and MHD.

Choose a background constant state and consider only perturbations in the $x$-direction. If we write the time evolution equations in terms of a matrix $A$ depending on the background and the vector of primitives $\altvec{P}$,
\begin{equation}
  \pp_t \altvec{P} + A \pp_x \altvec{P} = 0,
\end{equation}
then the linear waves are perturbations $\delta\altvec{P}$ that are eigenvectors of $A$.

In order to quantify the error, we evolve a wave for exactly one period and calculate the error for each primitive quantity $q$ as the $L^1$ norm of the difference between the initial and final states:
\begin{equation} \label{eq:l1_error}
  \epsilon_q = \frac{1}{N} \sum_{i=1}^N \abs{q^\mathrm{init}_i-q^\mathrm{fin}_i}.
\end{equation}
We then take the overall error to be the RMS of the set of errors $\epsilon_q$.

\begin{figure}
  \centering
  \includegraphics[width=\textwidth]{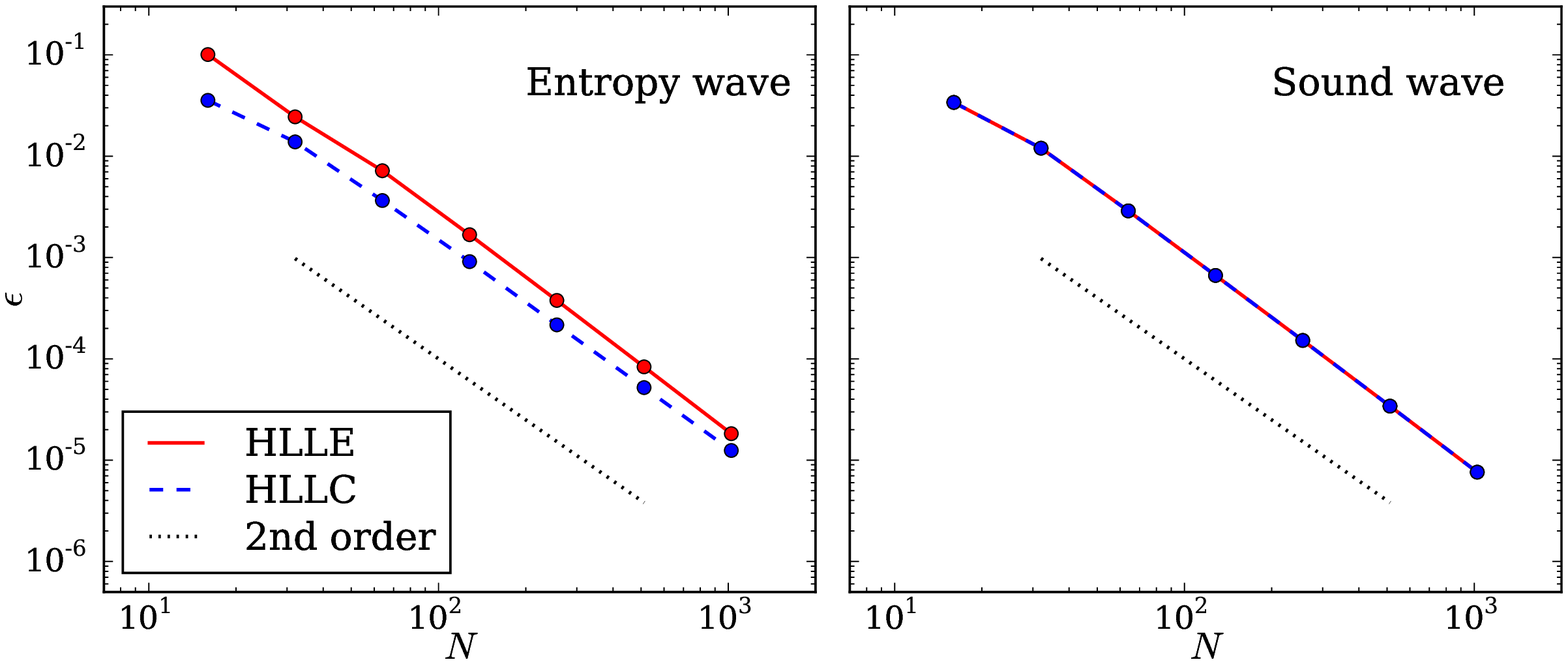}
  \caption{Convergence tests for special relativistic hydrodynamics. Both Riemann solvers converge at second order, with HLLC having a smaller overall error compared to HLLE for the entropy wave. LLF results are indistinguishable from HLLE. \label{fig:linear_sr_hydro}}
\end{figure}

The eigenvectors for hydrodynamics are given in \citet[their Appendix~A]{Falle1996}. We choose a background state of $\rho = 4$, $\pgas = 1$, $v^i = (0.1, 0.3, -0.05)$, with ratio of specific heats $\Gamma = 4/3$. The resulting wavespeeds are approximately $-0.30$, $0.1$, and $0.47$. We evolve the entropy and rightgoing sound wave, using both HLLE and HLLC Riemann solvers. The results are shown in Figure~\ref{fig:linear_sr_hydro}. Both methods converge at second order in the number of cells, as is expected since the reconstruction and time integration are both second order. Note that the error in the entropy wave is $1.7$ times larger using HLLE compared to HLLC for these parameters, while the solvers have the same error for the sound wave. This reflects the fact that when inside the wavefan, using just the sound speeds leads to averaging over too large a domain of dependence, diffusing the solution.

\begin{figure}
  \centering
  \includegraphics[width=\textwidth]{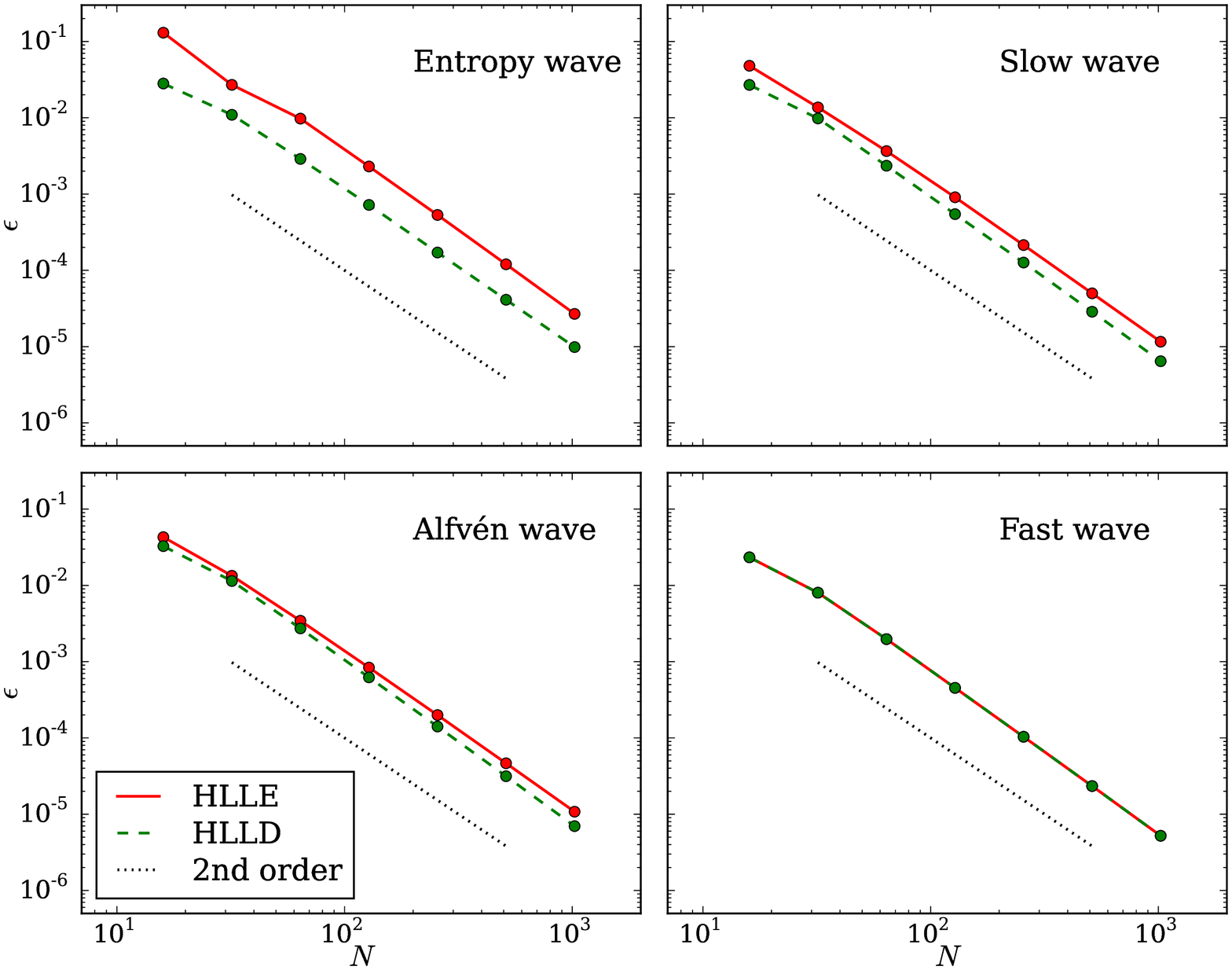}
  \caption{Convergence tests for special relativistic MHD. Both Riemann solvers converge at second order, with HLLD having a smaller overall error compared to HLLE for all but the fast wave. LLF results are indistinguishable from HLLE. \label{fig:linear_sr_mhd}}
\end{figure}

For MHD we take the same background state and add a magnetic field $B^i = (2.5, 1.8, -1.2)$, which results in all wavespeeds (approximately $-0.71$, $-0.55$, $-0.21$, $0.1$, $0.35$, $0.54$, and $0.81$) being well separated. The eigenvectors are given in \citet{Anton2010} as their \extref{46} (entropy), \extref{65} (Alfv\'en), and \extref{71} (magnetosonic). The results for the four rightmost waves are given in Figure~\ref{fig:linear_sr_mhd}. Again we have second order convergence, and again the more advanced Riemann solver has smaller errors for the internal waves of the Riemann fan. Here HLLD's error is a factor of $3.1$, $1.7$, and $1.4$ times lower than that of HLLE for the entropy, slow magnetosonic, and Alfv\'en waves, respectively.

We can also perform these same tests in a way that exercises the general relativistic portions of the code. Consider the ``tilted'' coordinates $(t', x', y', z')$ related to Minkowski coordinates $(t, x, y, z)$ by
\begin{subequations} \begin{align}
  t' & = \frac{t+ax}{\sqrt{1+a^2}}, \\
  x' & = \frac{x-at}{\sqrt{1+a^2}}, \\
  y' & = y, \\
  z' & = z
\end{align} \end{subequations}
for some parameter $a$, $\abs{a} < 1$. This coordinate system has its time axis corresponding to an observer moving with velocity $a$ with respect to the Minkowski frame, while the surface of constant time corresponds to that of an observer with velocity $-a$. Visually this corresponds to rigidly rotating the $tx$-axes into the $t'x'$-axes. While the connection coefficients and thus source terms vanish in this coordinate system, the $t'x'$ metric coefficients are nonvanishing and so there are off-diagonal components in the transformation matrices \eqref{eq:to_global} and \eqref{eq:to_local}. Additionally, the interface velocity \eqref{eq:interface_velocity} is nonzero.

We can rerun the previous convergence tests in this new coordinate system. We use the same background states, with the given numbers understood to be in Minkowski coordinates. The linear solution is known throughout spacetime, and we initialize the perturbation over a single wavelength on the surface $t' = 0$. This is evolved until $t' = (1 + a\lambda) / \lvert \lambda-a \rvert$, which corresponds to one wavelength crossing the domain at Minkowski $3$-velocity $\lambda$. For most waves we choose $a = 0.1$. For the entropy waves we choose $a = 0.05$, since if $a$ matches $\lambda$ the HLLC and HLLD solvers will exactly capture the wave, placing our comparison \eqref{eq:l1_error} at the machine precision noise floor even for low resolutions.

\begin{figure}
  \centering
  \includegraphics[width=\textwidth]{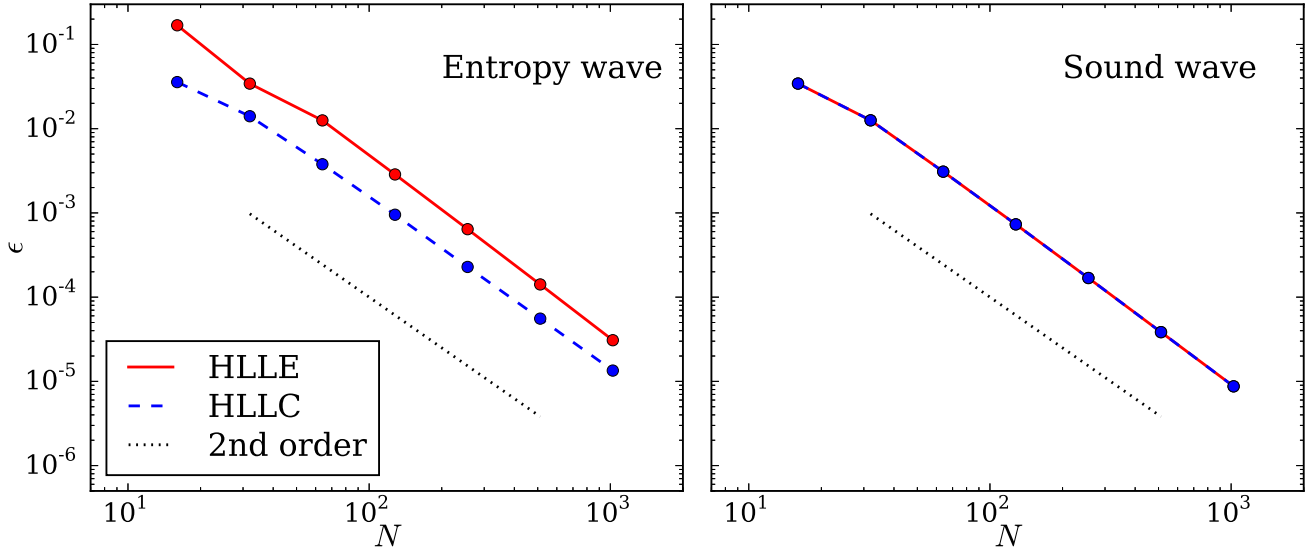}
  \caption{Convergence tests for general relativistic hydrodynamics. Both Riemann solvers converge at second order, with HLLC improving upon HLLE for the entropy wave, as in Figure~\ref{fig:linear_sr_hydro}. \label{fig:linear_gr_hydro}}
\end{figure}

\begin{figure}
  \centering
  \includegraphics[width=\textwidth]{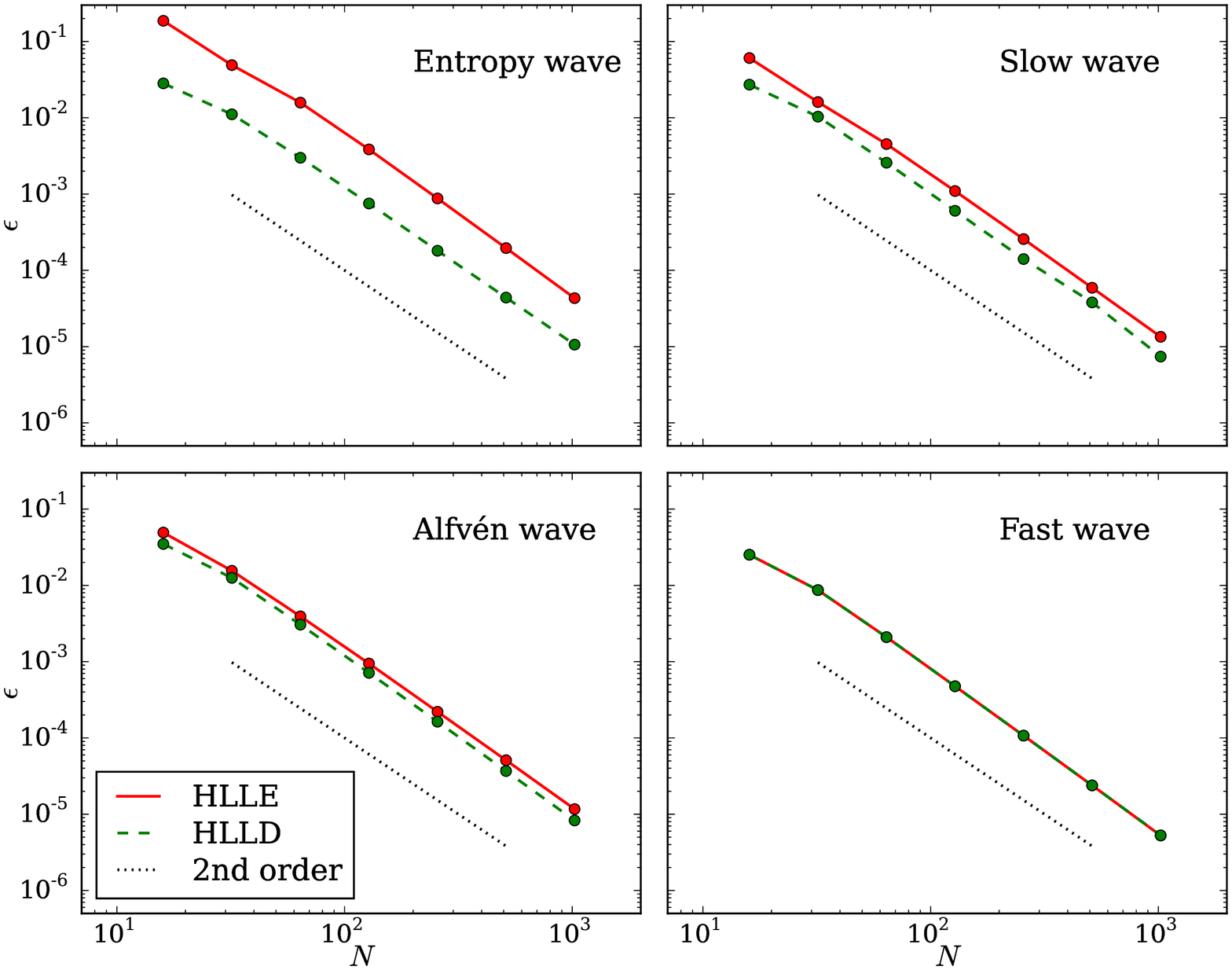}
  \caption{Convergence tests for general relativistic MHD. Both Riemann solvers converge at second order, with HLLD improving upon HLLE for all but the fast wave, as in Figure~\ref{fig:linear_sr_mhd}. \label{fig:linear_gr_mhd}}
\end{figure}

The results for hydrodynamics and MHD are given in Figures~\ref{fig:linear_gr_hydro} and~\ref{fig:linear_gr_mhd}. For the hydrodynamic entropy wave, the HLLC solver's error is lower than that of HLLE by a factor of $2.8$. In MHD, the HLLD solver's error beats that of HLLE by $4.8$, $1.8$, and $1.4$ in the entropy, slow, and Alfv\'en waves. We note that in these particular tests the advanced Riemann solvers outperform HLLE by larger margins in general relativity compared to special relativity.

\subsection{Shock Tubes}
\label{sec:tests:shocks}

Another class of one-dimensional special relativistic tests is that of shock tubes. While the above wave tests focus on linear perturbations, shock tubes emphasize the nonlinear behavior of the equations, as well as the ability to resolve discontinuities in the solutions.

\begin{figure}
  \centering
  \includegraphics[width=\textwidth]{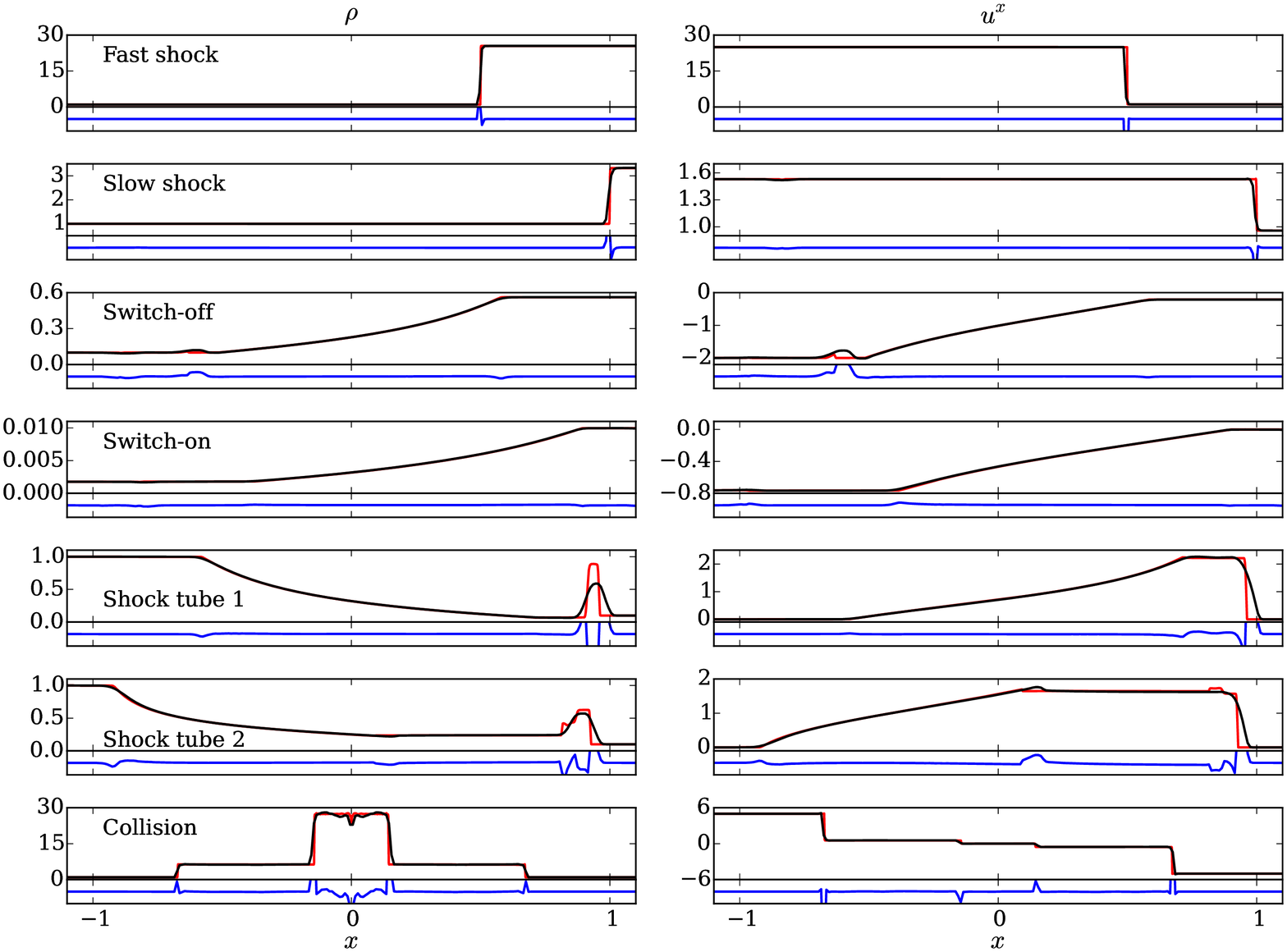}
  \caption{\Citeauthor{Komissarov1999}\ shock tubes for special relativistic MHD. Results are computed with HLLD on a grid of $220$ (black) and $2200$ (red) cells. The blue lines show the relative error between the two resolutions normalized by the maximum value attained by the variable over the domain, with the plot limits being $\pm10\%$ in all cases. \label{fig:nonlinear_waves}}
\end{figure}

We show the results for seven shock tubes from \citet{Komissarov1999} \citep[cf.][their Figures~\extref{4} and \extref{5}]{Gammie2003} in Figure~\ref{fig:nonlinear_waves}. These results in black all use the HLLD solver with $220$ grid cells in the interval $[-1.1,1.1]$, corresponding to the same fixed resolution used by \citeauthor{Gammie2003} in all cases. The underplotted red line is a reference solution computed with ten times the resolution. The CFL number is $0.8$ except in the cases of the fast shock ($0.2$) and the collision ($0.5$).

Some of the same numerical artifacts observed by \citeauthor{Komissarov1999}\ can be seen in our results. For example, we see ringing upstream of the slow shock, a spurious bump at the low-density end of the switch-off rarefaction wave, and overshooting at the low-density end of the shock tube~2 rarefaction wave. In the case of shock tube~1 we do not observe the overshooting in density at the base of the density rarefaction wave, and we have greatly diminished overshooting at the top of the velocity discontinuity, despite using the same resolution as \citeauthor{Komissarov1999}. That HLLD in particular does well at capturing shocks is not surprising; \citet{Mignone2009} note this in comparing different Riemann solvers. We conclude that we are correctly capturing the nonlinear behavior expected in this special relativistic test suite.

\subsection{Spherical Blast Wave}
\label{sec:tests:blast}

As a two-dimensional test of our general relativistic formulation, we evolve a spherical blast wave. The initial conditions are uniform density ($\rho = 1$) with a central overpressure ($\pgas = 2.5$ inside a radius of $0.5$; $\pgas = 0.1$ outside). We enforce periodic boundary conditions on a rectangular grid of width $4$ and height $6$. By time $t = 15$, the blast wave has intersected itself a number of times, inducing Richtmyer-Meshkov instability features. The shock velocities here are not so high as to be in the regime where Richtmyer-Meshkov instability is strongly suppressed, as discussed in \citet{Mohseni2014} and \citet{Zanotti2015}.

In order to test nontrivial spatial geometry, we evolve the system not only in Minkowski coordinates but also in ``snake'' coordinates that vary sinusoidally with Minkowski position. Explicitly, given Minkowski $(t, x, y, z)$, define
\begin{subequations} \begin{align}
  x^0 & = t, \\
  x^1 & = x, \\
  x^2 & = y + a \sin kx, \\
  x^3 & = z,
\end{align} \end{subequations}
where $a$ and $k$ are free parameters. If we define the parameter
\begin{equation}
  \delta = a k \cos kx^1,
\end{equation}
the metric becomes
\begin{equation}
  g_{\mu\nu} =
  \begin{pmatrix}
    -1 & 0                 & 0       & 0 \\
    0  & \sqrt{1+\delta^2} & -\delta & 0 \\
    0  & -\delta           & 1       & 0 \\
    0  & 0                 & 0       & 1
  \end{pmatrix}.
\end{equation}
The transformations \eqref{eq:to_global} and \eqref{eq:to_local} differ from the identity via $\tensor{M}{^2_{\hat{x}}} = -\tensor{M}{^{\hat{y}}_1} = \delta$, and there is a source term due to $\Gamma^2_{11}$.

\begin{figure}
  \centering
  \includegraphics[width=\textwidth]{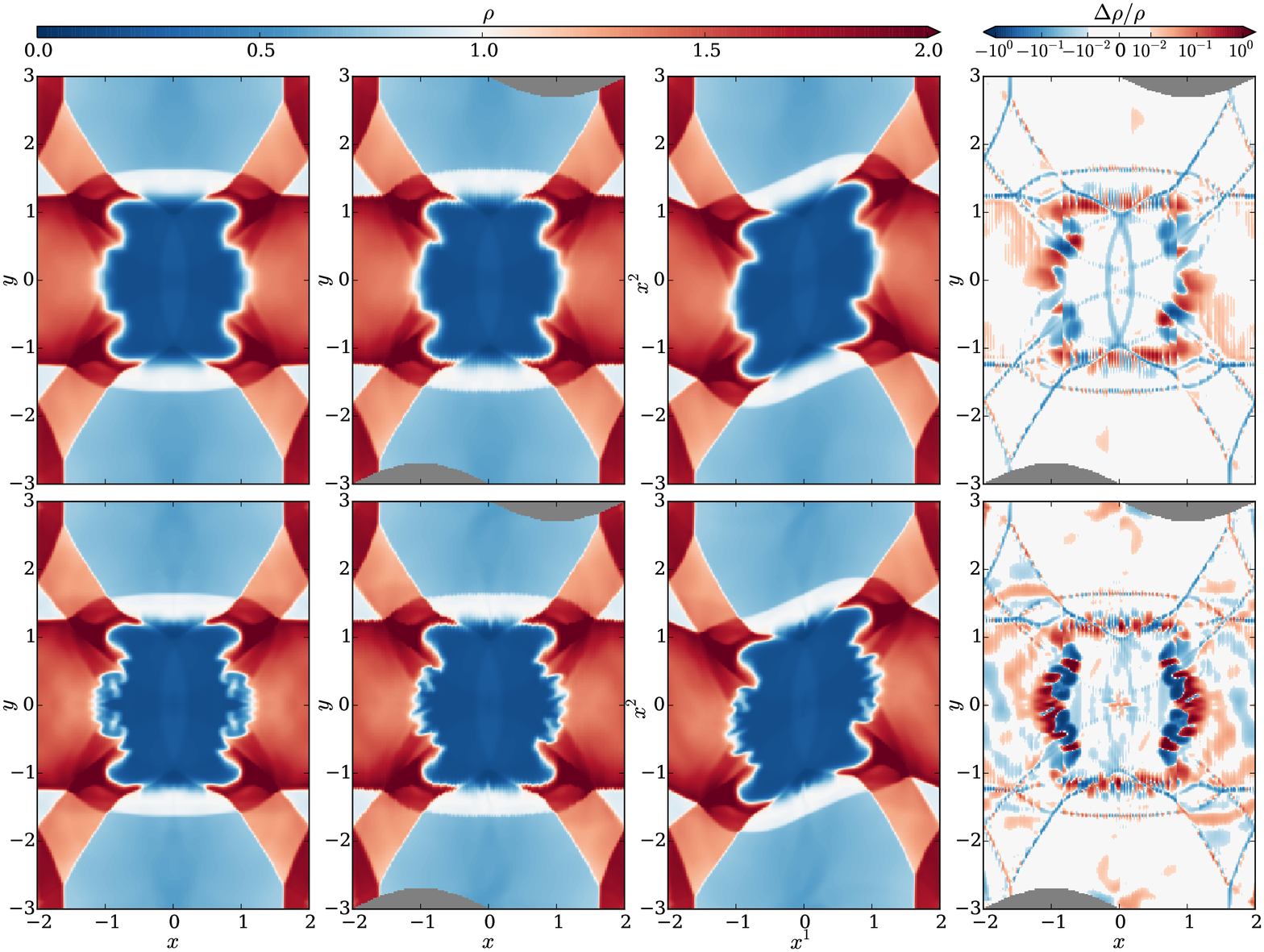}
  \caption{Density $\rho$ for the spherical blast wave at $t = 15$ in Minkowski coordinates (left), snake coordinates transformed back to Minkowski (center left), and snake coordinates (center right). The right panels show the fractional difference $\Delta\rho/\rho$ between the first and second columns, where $\Delta\rho = \rho_\mathrm{snake} - \rho_\mathrm{Minkowski}$ and white regions have a fractional difference of less than $1\%$. The top panels use HLLE, while the bottom use HLLC. \label{fig:blast_hydro}}
\end{figure}

The hydrodynamical blast wave density is shown in Figure~\ref{fig:blast_hydro}. The top row displays results using the HLLE Riemann solver, while the bottom row uses HLLC. The first column shows the results at $t = 15$ as computed in Minkowski coordinates on a $200\times300$ grid with a CFL number of $0.4$. The second column shows the same results as calculated in snake coordinates and transformed back to Minkowski at the end of the calculation. Here we also use a $200\times300$ grid, with $a = 0.3$ and $k = \pi/2$. Lines of constant $x^2$ in snake coordinates follow the sinusoidal variation of the shaded region.

One can see that the solutions remain qualitatively unchanged when the coordinates are varied, despite the fact that the fluxes, the source terms, and even the conserved quantities themselves vary between coordinate systems. To give a sense of how the internal representation of the fluid changes between coordinate systems, the third column of Figure~\ref{fig:blast_hydro} plots density in snake coordinates $(x^1,x^2)$.

For a more quantitative comparison, the last column shows the fractional difference in density between the Minkowski solution and the snake solution transformed to Minkowski coordinates. Colored areas indicate regions where the fractional difference exceeds $1\%$. Differences are generally confined to cell-level variations in the locations of shock fronts, or else to the details of the Richtmyer-Meshkov fingering. This latter effect is to be expected with an instability seeded with grid noise. We also note that the HLLC Riemann solver leads to more sharply defined fingers, whereas using just HLLE the structures are notably more diffuse.

As a further test, we add an initially uniform magnetic field ($B^i = (1, 1, 0)$) and rerun the blast wave, this time with the HLLD solver. In this strongly magnetized regime, the blast wave should be guided by the magnetic field lines, breaking its initial symmetry.

\begin{figure}
  \centering
  \includegraphics[width=\textwidth]{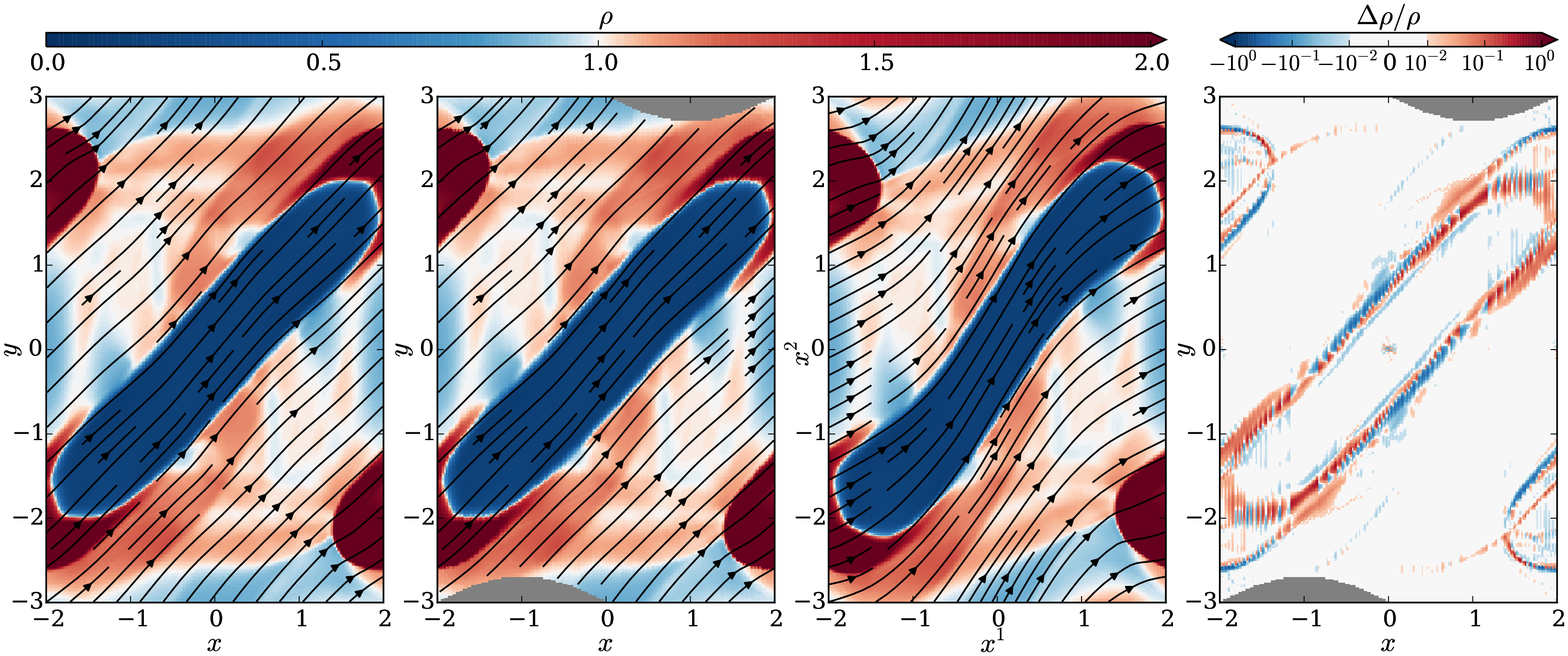}
  \caption{Similar to Figure~\ref{fig:blast_hydro} with MHD added. Density $\rho$ (color) and magnetic field $B^i$ (streamlines) are shown for the strongly magnetized blast wave at $t = 5$ in Minkowski coordinates (left), snake coordinates transformed back to Minkowski (center left), and snake coordinates (center right). The right panel shows the fractional difference between the first and second columns as in Figure~\ref{fig:blast_hydro}. All panels use HLLD. \label{fig:blast_mhd}}
\end{figure}

The MHD results are shown in Figure~\ref{fig:blast_mhd}, with color denoting $\rho$ and streamlines showing the magnetic field. The same layout is used here as in both rows of Figure~\ref{fig:blast_hydro}. Both MHD calculations are done with HLLD on a $200\times300$ periodic grid with a CFL number of $0.4$.

As expected, the field guides the blast wave and suppress the Richtmeyer-Meshkov instability. Moreover, the same results are found in both coordinate systems. The self-interaction of the wave after wrapping around the boundaries induces a slight curvature in the boundary of the rarefaction region and in the magnetic field lines, and even this is the same in the two coordinate systems.

\subsection{Bondi Accretion}
\label{sec:tests:bondi}

As a test of the code in a nontrivial spacetime, we model Bondi accretion onto a Schwarzschild black hole. That is, we seek steady-state solutions to radial flow onto a black hole, where pressure can be nonnegligible.

The Bondi solution is derived in \citet{Hawley1984}, and we summarize the pertinent formulas here. Begin with a black hole mass $M$, an adiabatic index $\Gamma$, an adiabat $K$, and a critical radius $\rcrit$. Define the polytropic index $n = 1/(\Gamma-1)$. The critical velocity and temperature are then
\begin{gather}
  \ucrit^1 = -\sqrt{\frac{M}{2\rcrit}}, \\
  \Tcrit = \frac{n}{n+1} \cdot \frac{\ucrit^1\ucrit^1}{1-(n+3)\ucrit^1\ucrit^1}.
\end{gather}
Next we calculate the constants
\begin{align}
  C_1 & = \Tcrit^n \ucrit^1 \rcrit^2, \\
  C_2 & = \big(1 + (n+1)\Tcrit\big)^2 \left(1 - \frac{2M}{\rcrit} + \ucrit^2\ucrit^1\right).
\end{align}
The run of temperature with radius is given by solving the equation
\begin{equation}
  \big(1 + (n+1)T\big)^2 \left(1 - \frac{2M}{r} + \frac{C_1^2}{r^4T^{2n}}\right) = C_2.
\end{equation}
There will generally be two roots to this equation. The lesser root is taken for $r < \rcrit$, while the greater root is taken outside the critical point. We can then find the radial velocity according to
\begin{equation}
  u^1 = \frac{C_1}{r^2T^n}.
\end{equation}
Finally, density and pressure are given by
\begin{align}
  \rho & = \left(\frac{T}{K}\right)^n, \\
  \pgas & = T \rho.
\end{align}
Note these formulas assume Schwarzschild coordinates.

We choose parameters $M = 1$, $\Gamma = 4/3$, $K = 1$, and $\rcrit = 8$, and we simulate the region $3 < r < 10$, $\pi/4 < \theta < 3\pi/4$. In order to test MHD as well as pure hydrodynamics, we optionally include a purely radial monopole field with $B^r \propto 1/r^2$, normalized such that $b_\lambda b^\lambda / \rho = 10$ at the inner radius. This corresponds to plasma $\beta = 0.246$ at the critical point. The 2D domain is divided into $N$ equally spaced cells in each direction. We initialize the Bondi solution and run the simulation until time $t = 10$, computing the $L^1$ difference in $\pgas$ between the beginning and end. We use the inner $3/4$ of the grid in each direction, and normalize the errors $\epsilon$ by the $L^1$ norm of the initial $\pgas$.

\begin{figure}
  \centering
  \includegraphics[width=6in]{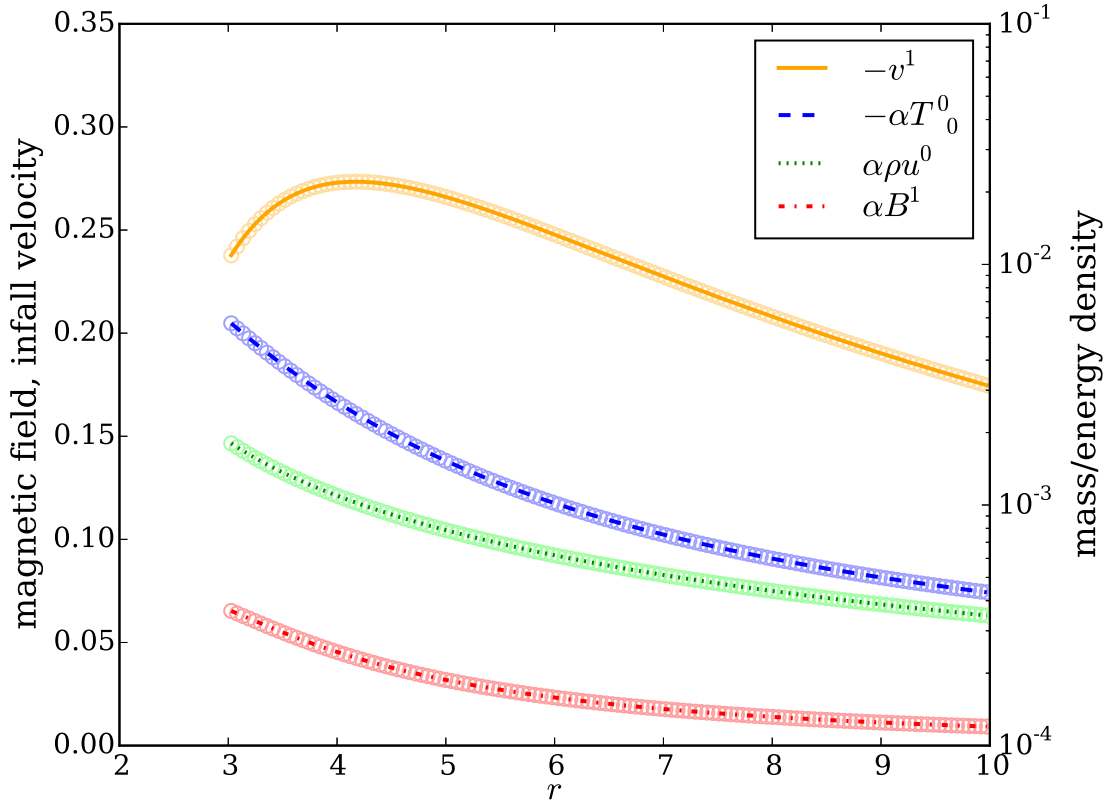}
  \caption{Magnetized Bondi flow in Schwarzschild coordinates with parameters as described in the text. Lines indicate the initial solution, while the circles indicate the values at each cell at $t = 10$. The conserved variables are scaled by the lapse $\alpha$. \label{fig:bondi_solution}}
\end{figure}

\begin{figure}
  \centering
  \includegraphics[width=0.45\textwidth]{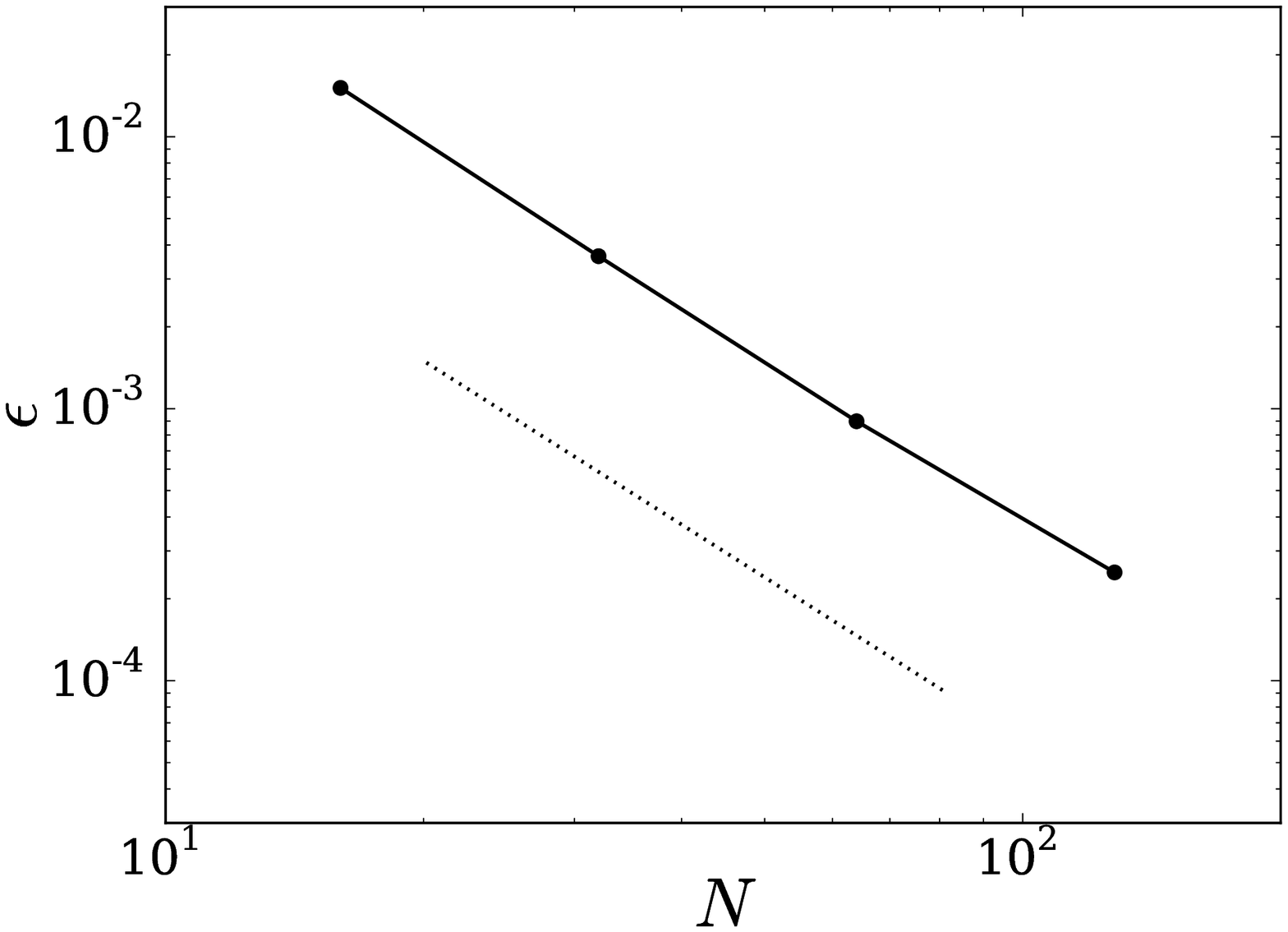}
  \includegraphics[width=0.45\textwidth]{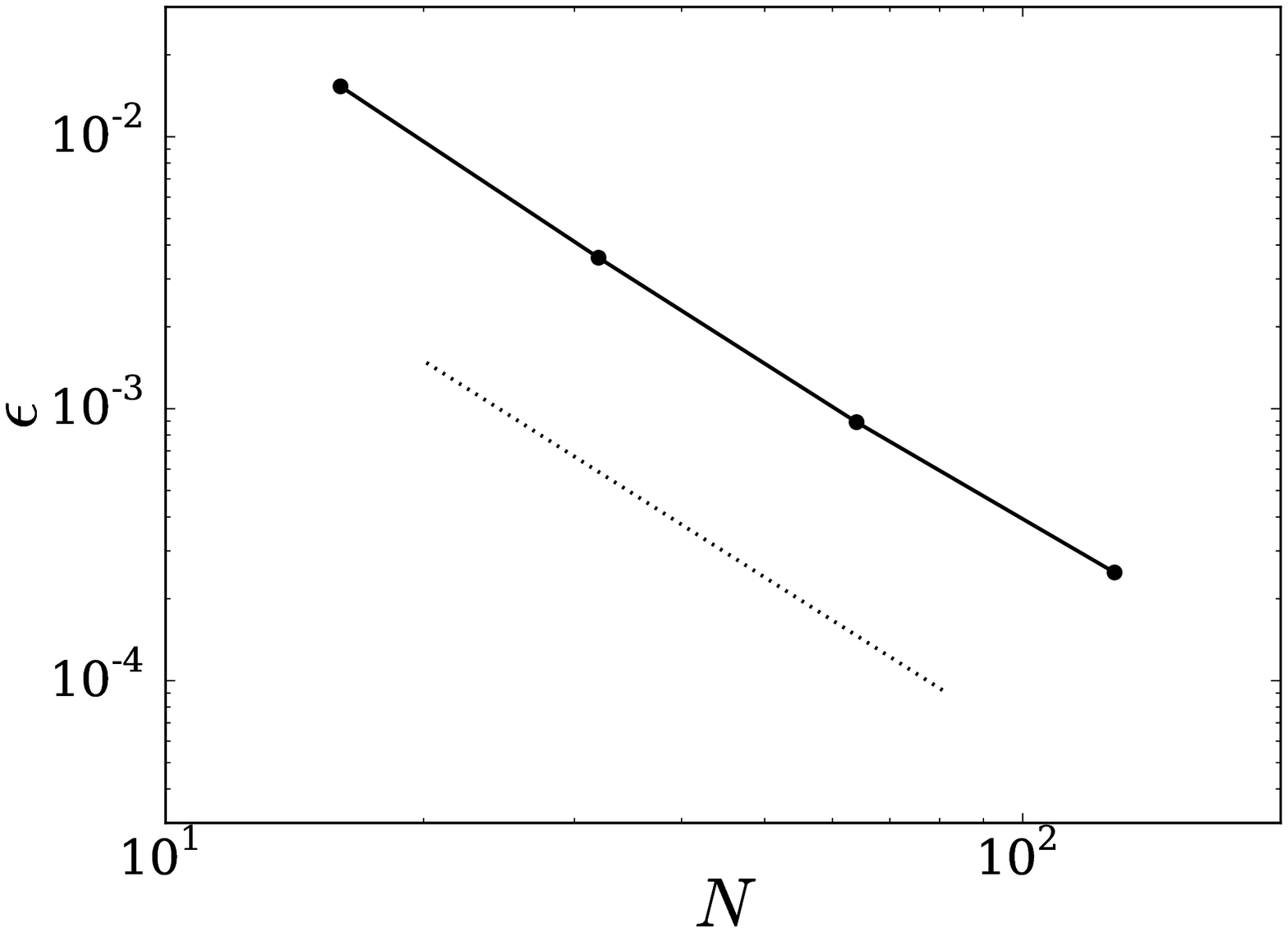}
  \caption{Errors in gas pressure for hydrodynamical (left) and MHD (right) Bondi flow on various $N \times N$ grids. In both cases we achieve second-order convergence, as indicated by the dotted lines. \label{fig:bondi_error}}
\end{figure}

The steady-state solution is shown in Figure~\ref{fig:bondi_solution}. The convergence is second order for both hydrodynamics using HLLC and MHD using HLLD, as shown in Figure~\ref{fig:bondi_error}. Since this is a stationary problem, convergence is not affected by the order of the integrator. However, it is limited by the order of reconstruction, and so these tests show we are correctly getting the full second-order convergence in space expected, even in a curved spacetime.

\subsection{Magnetized Equatorial Inflow}
\label{sec:tests:inflow}

For a final test problem, we simulate the plunging region of a magnetized thin disk around a spinning black hole, as described in \citet{Gammie1999}. We choose the same values for the free parameters as in \citet{Gammie2003}:\ $M = 1$, $a/M = 0.5$, $F_M = -1$, and $F_{\theta\phi} = 0.5$, with the flow matching onto the innermost stable circular orbit. The simulation covers the range $2 < r < 4$ (where the horizon is at $r = 1.86603$ and the innermost stable circular orbit is at $r = 4.23300$). We use the same Kerr-Schild coordinates $\{t, r, \theta, \phi\}$ as \citeauthor{Gammie2003}, with line element
\begin{equation} \begin{split}
  ds^2 & = -\left(1-\frac{2Mr}{\Sigma}\right) \, \dt^2 + \frac{4Mr}{\Sigma} \, \dt\,\dr - \frac{4Mar}{\Sigma} \sin^2\!\theta \, \dt\,\dph + \left(1+\frac{2Mr}{\Sigma}\right) \, \dr^2 \\
  & \quad - 2a \left(1+\frac{2Mr}{\Sigma}\right) \sin^2\!\theta \, \dr\,\dph + \Sigma \, \dth^2 + \left(r^2+a^2+\frac{2Ma^2r}{\Sigma}\sin^2\!\theta\right) \sin^2\!\theta \, \dph^2.
\end{split} \end{equation}
Here $\Sigma = r^2 + a^2 \cos^2\!\theta$. Note that Kerr-Schild $r$ and $\theta$ are the same as Boyer-Lindquist $r$ and $\theta$.

\begin{figure}
  \centering
  \includegraphics[width=6in]{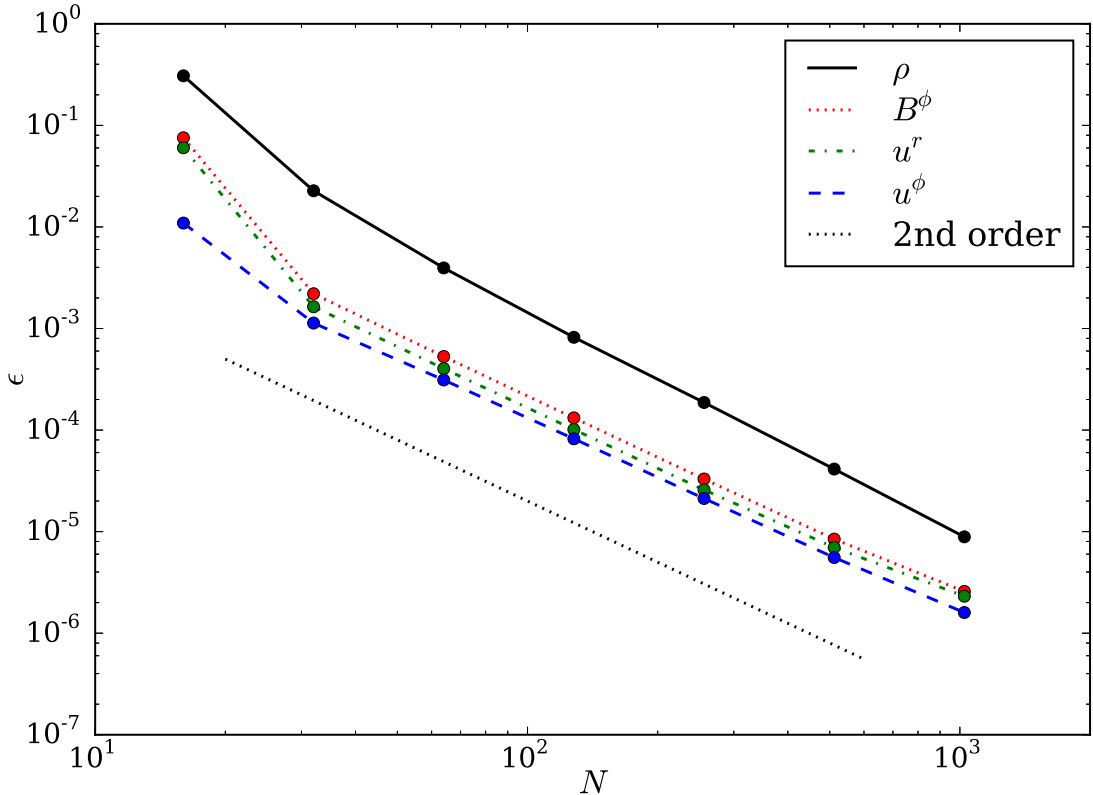}
  \caption{Errors in density, azimuthal magnetic field, and radial and azimuthal velocities as functions of the number of radial grid cells in the magnetized inflow test. As expected, all quantities converge at second order. \label{fig:magnetized_inflow}}
\end{figure}

The flow is initialized to the steady-state solution, with gas pressure set to $10^{-10}$ times the density, and is run to time $t = 15$. A plot of the $L^1$ error in four quantities is shown in Figure~\ref{fig:magnetized_inflow}. We find second-order convergence, demonstrating that the code is solving the correct equations even in a highly nontrivial spacetime. In particular, this test is performed with a frame-transforming HLLE Riemann solver, showing that the frame transformations work even when there is a large shift between the frames.

\section{Torus Simulations}
\label{sec:torus}

As a full demonstration of the code in a science application setting, we simulate a fluid torus around a spinning black hole. As proven in \citet{Fishbone1976}, there are steady-state solutions for hydrodyamical tori of an isentropic fluid with constant angular momentum orbiting in Kerr spacetime. Given a black hole mass and spin and a fluid equation of state, the solutions are defined by choosing the radius of the inner edge of the torus, $\rin$, and either the radius of the pressure maximum, $\rpeak$, or the angular momentum per unit mass, $\ell = u^t u_\phi$ (in Boyer-Lindquist coordinates).

A general relativistic code should be able to maintain such a torus, balancing gravity, centrifugal force, and pressure gradients. We take a black hole with mass $M = 1$ and spin $a/M = 0.95$ and initialize a $\Gamma = 13/9$ fluid torus with $\rin = 3.7$ and $\ell = 3.85$ (so $\rpeak = 7.82$). The torus is evolved on a 2D grid with $N_r$ cells in the radial direction and $N_r/2$ cells in the polar direction using the aforementioned Kerr-Schild coordinates. Our grid extends radially from $0.98$ times the outer horizon radius $M + \sqrt{M^2-a^2}$ to $r = 20$, with geometric grid spacing such that each cell is $1.025$ times as wide as its inner neighbor. We limit the polar angle to $\pi/4 < \theta < 3\pi/4$, excising parts of the necessarily non-stationary atmosphere that should not impact the torus proper. This simulation employs uniform spacing in $\theta$. We impose outflow conditions on the inner and outer boundaries and reflecting boundary conditions on the constant-$\theta$ boundaries.

\begin{figure}
  \centering
  \includegraphics[width=0.45\textwidth]{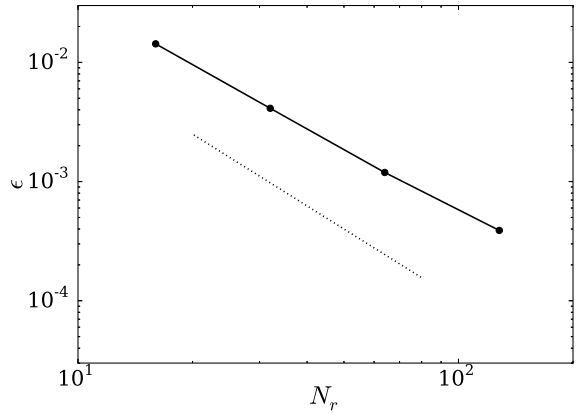}
  \caption{Errors in density for 2D Fishbone-Moncrief hydrodynamical tori on various $N_r \times N_r/2$ grids. Errors are defined by \eqref{eq:torus_error}. The dotted line indicates second-order convergence. \label{fig:torus_hydro_convergence}}
\end{figure}

The simulation is evolved using the HLLC solver with a CFL number of $0.2$ until a time $t = 1$, at which point the error accrued in $\rho$ is calculated. Define the region $T$ to be those cells for which $\rho$ is initially at least $0.02$ its peak value. We calculate the error $\epsilon$ over the region $T$ (thus avoiding edge effects) in the $L^1$ sense:
\begin{equation} \label{eq:torus_error}
  \epsilon = \frac{\displaystyle\int_T\abs{\rho_\mathrm{fin}-\rho_\mathrm{init}}\sqrt{-g}\,\dr\,\dth}{\displaystyle\int_T\abs{\rho_\mathrm{init}}\sqrt{-g}\,\dr\,\dth}.
\end{equation}
The errors are shown in Figure~\ref{fig:torus_hydro_convergence}, where one can see that they are second order. This is the same test problem as done in \citet[cf.\ their Figure~15]{Gammie2003} with the exception that we use a slightly different grid with no equatorial compression of the constant-$\theta$ surfaces.

Our torus is indeed able to maintain equilibrium, even over longer times. To demonstrate this we continue the $64\times32$ simulation to $t = 430$, corresponding to three orbital periods at the pressure maximum, using both HLLE and HLLC solvers. With the HLLE solver the error is $\epsilon = 0.030$, while it is $\epsilon = 0.021$ with HLLC.

We next consider a 3D Fishbone-Moncrief torus with an added magnetic field, upsetting equilibrium via the magnetorotational instability (MRI). Here we choose $M = 1$, $a/M = 0.9375$, $\Gamma = 13/9$, $\rin = 6$, and $\rpeak = 12$ (so $\ell = 4.28$). The grid is expanded to $0.98 (M + \sqrt{M^2-a^2}) < r < 40$ and $0 < \theta < \pi$ (with no artificial boundary at the poles), and it covers the entire azimuthal range. In one case we use the equivalent of $144\times128\times96$ cells in the radial, polar, and azimuthal directions, with geometric radial spacing (ratio $1.02397$) and uniform spacing in the angles. Static mesh refinement is used to derefine the polar regions outside the torus by a factor of $4$ in each dimension, thus preventing the timestep from being limited by those regions. Explicitly, there are $80$ cells equally spaced in the range $3\pi/16 < \theta < 13\pi/16$. For a higher resolution grid, we further double the number of cells in each dimension in the region $7\pi/32 < \theta < 25\pi/32$. This forms an effective $288\times256\times192$ grid.

To this torus we add a magnetic field with purely azimuthal vector potential $A^\phi \propto \max(\rho-0.2, 0)$, calculated from a density normalized to unity at its maximum. The magnetic field is normalized such that the ratio of maximum gas pressure to maximum magnetic pressure (necessarily located at different points) is $100$. These parameters are chosen to roughly match those of \citet{Shiokawa2012} except in not having an equatorially compressed grid like theirs. In particular, our low and high resolution runs have $\Delta\theta = 0.0245$ and $\Delta\theta = 0.0123$ in their respective most refined regions, while the $96\times96\times64$ and $144\times144\times96$ grids from \citeauthor{Shiokawa2012}\ have $\Delta\theta = 0.00983$ and $\Delta\theta = 0.0065$ at the midplane.

\begin{figure}
  \centering
  \includegraphics[width=\textwidth]{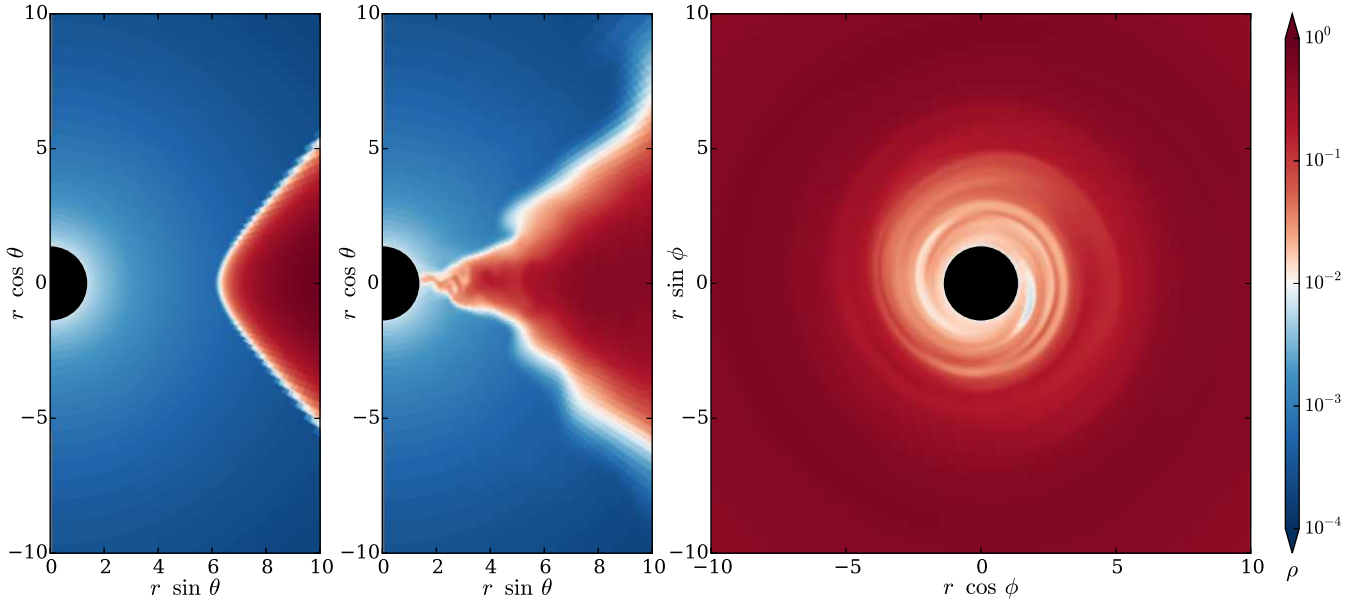}
  \caption{Density $\rho$ of a 3D Fishbone-Moncrief MHD torus with effective resolution $288\times256\times192$. Shown are the initial conditions in the $r\theta$-plane (left), the final state at $t = 12{,}000$ in the same plane (center), and the final state in the $r\phi$-plane (right). \label{fig:torus_slices}}
\end{figure}

We evolve the system using an LLF Riemann solver just as is used in Harm. The simulation is run until a time $t = 12{,}000$. For comparison, a geodesic circular orbit at the pressure maximum has a period of $t = 267$. After several orbits the MRI sets in, causing turbulence that eventually disrupts the entire torus. The initial and final densities for the run are illustrated in Figure~\ref{fig:torus_slices}.

\begin{figure}
  \centering
  \includegraphics[width=\textwidth]{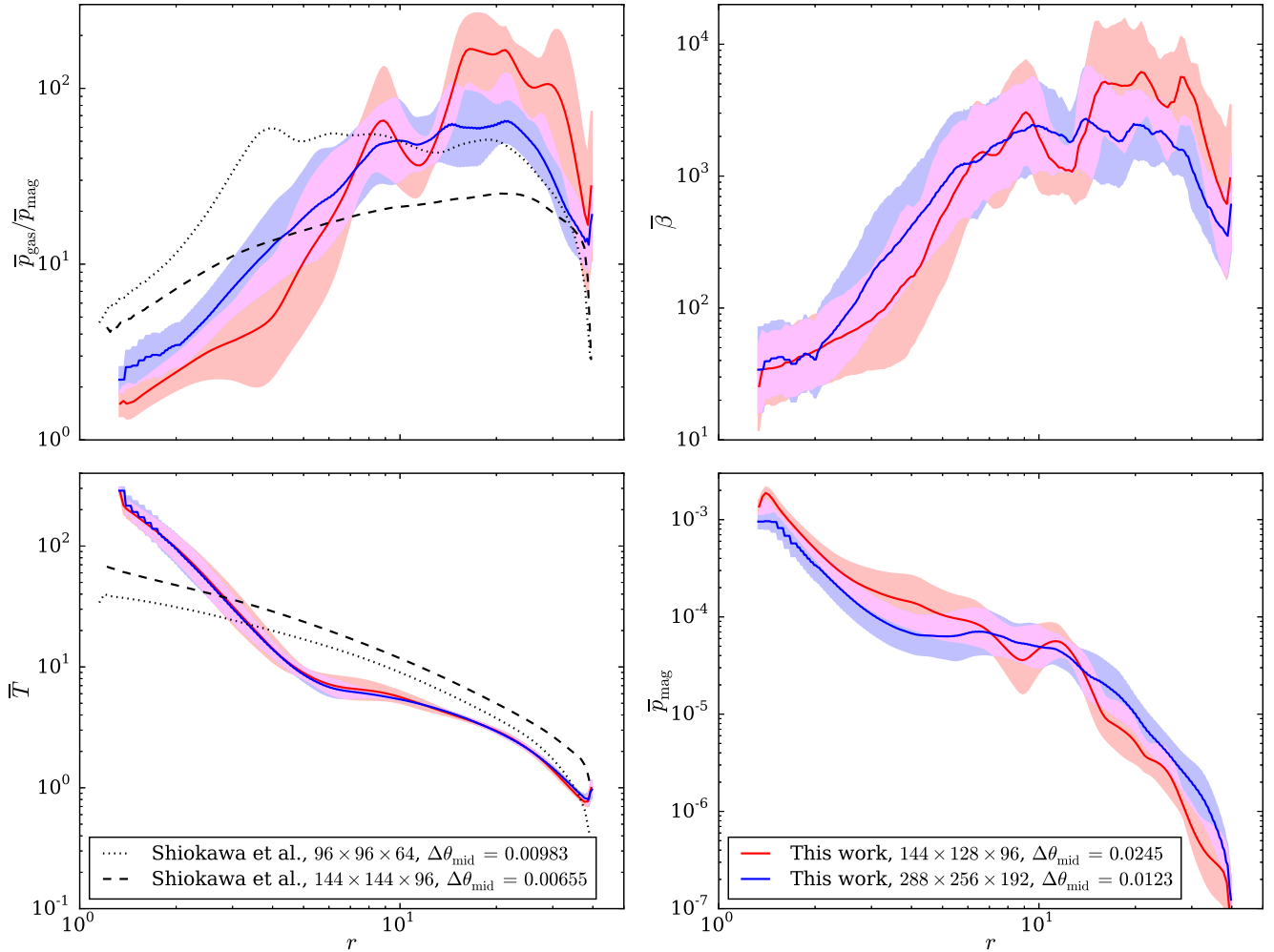}
  \caption{Spherically averaged, time-averaged profiles of various quantities in the 3D torus. These are the gas-to-magnetic pressure ratio (spherical averages taken before ratio, time averages take after; top left), plasma $\beta$ (top right), electron temperature (bottom left), and magnetic pressure (bottom right). The left two panels show the same quantities as plotted in Figure~2 of \citet{Shiokawa2012}, corresponding to their ``soft'' polar boundary condition. In all cases shaded regions indicate $1\sigma$ standard deviations from $800$ samples at different times. \label{fig:torus_profile}}
\end{figure}

For a more quantitative analysis, we construct radial profiles of physical quantities. Following \citet[cf.\ their \extref{9} and \extref{10}]{Shiokawa2012}, we define a spherical shell average of a quantity $q$ at a Kerr-Schild time $t$ according to
\begin{equation}
  \bar{q} = \frac{\displaystyle\int_0^{2\pi}\!\!\int_{\theta_\mathrm{min}}^{\theta_\mathrm{max}}q\rho\sqrt{-g}\,\dth\,\dph}{\displaystyle\int_0^{2\pi}\!\!\int_{\theta_\mathrm{min}}^{\theta_\mathrm{max}}\rho\sqrt{-g}\,\dth\,\dph},
\end{equation}
where in our case we take $(\theta_\mathrm{min},\theta_\mathrm{max}) = (\pi/4,3\pi/4)$ in order to exclude any effects of the atmosphere. We then average $\bar{q}$ over the time range $4000$ to $12{,}000$, noting the mean and standard deviation of values at each radius. This is done for plasma $\beta$, electron temperature $T = \mprot\pgas/2\melec\rho$ (called $\theta_\mathrm{e}$ by \citeauthor{Shiokawa2012}), and magnetic pressure $\pmag = b^\lambda b_\lambda/2$ in Figure~\ref{fig:torus_profile}. For a more direct comparison to \citeauthor{Shiokawa2012}\ we also show the result of time-averaging $\bar{p}_\mathrm{gas}/\bar{p}_\mathrm{mag}$, the result of which is simply called $\beta$ in that work.

The electron temperature roughly matches what \citeauthor{Shiokawa2012}\ find, both in magnitude and in trend with radius, though they do not have the same kink at around $r = 5\text{--}6$ and our disk gets notably hotter inside the innermost stable circular orbit at $r = 2.04420$. We get the same shape as \citeauthor{Shiokawa2012}\ for $\pgas/\pmag$, especially their low resolution ``soft'' polar run (cf.\ the upper left panel of their Figure~2; data reproduced in Figure~\ref{fig:torus_profile}). They find $\pgas/\pmag$ to be less than about $60$ in all cases, whereas our simulations get up to about $200$ in the outer regions. Our ratio also approaches unity in the innermost regions, while theirs never drops below $4$. The same shape is found when we perform spherical averages of $\beta$ itself, rather than $\pgas$ and $\pmag$ separately, as shown in the upper right of Figure~\ref{fig:torus_profile}. While there are some differences, we also note that as we increase resolution our $\pgas/\pmag$ profile approaches theirs, and we expect even better agreement if we were to have midplane $\Delta\theta$ values as small as theirs.

\Citeauthor{Shiokawa2012}\ also examined azimuthal correlation lengths in the equatorial plane. We use the same definitions as they do (their \extref{13} through \extref{17}), which we summarize here. For a vector quantity with components $q^\mu$ define the time-dependent correlation function
\begin{equation}
  R(r,\phi,t) = \frac{1}{r\Delta r\Delta\theta} \int_{\theta_\mathrm{min}}^{\theta_\mathrm{max}} \int_{r_\mathrm{min}}^{r_\mathrm{max}} \int_0^{2\pi} \delta q^\mu(r',\theta',\phi_0,t) \delta q_\mu(r',\theta',\phi_0+\phi,t) \, \dph \, r' \, \dr'\,\dth'.
\end{equation}
The integrals run over a region of radial width $\Delta r = r_\mathrm{max} - r_\mathrm{min}$ of a single cell and polar width $\Delta \theta = \theta_\mathrm{max} - \theta_\mathrm{min}$ of two cells bordering the equatorial plane. The quantities $\delta q^\mu$ and $\delta q_\mu$ are deviations from averages of $q^\mu$ and $q_\mu$ over the domains of integration. For scalar quantities we simply use the integrand $\delta q(r',\theta',\phi_0,t) \delta q(r',\theta',\phi_0+\phi,t)$ instead. Define the time-averaged correlation function as
\begin{equation}
  \bar{R}(r,\phi) = \int_{t_\mathrm{min}}^{t_\mathrm{max}} \frac{R(r,\phi,t)}{R(r,0,t)} \, \dt,
\end{equation}
where we average from $4000$ to $12{,}000$. Then the correlation length $\lambda$ at a given radius $r$ is the value for which $\bar{R}(r,\lambda) = \bar{R}(r,0)/\ee$.

\begin{figure}
  \centering
  \includegraphics[width=\textwidth]{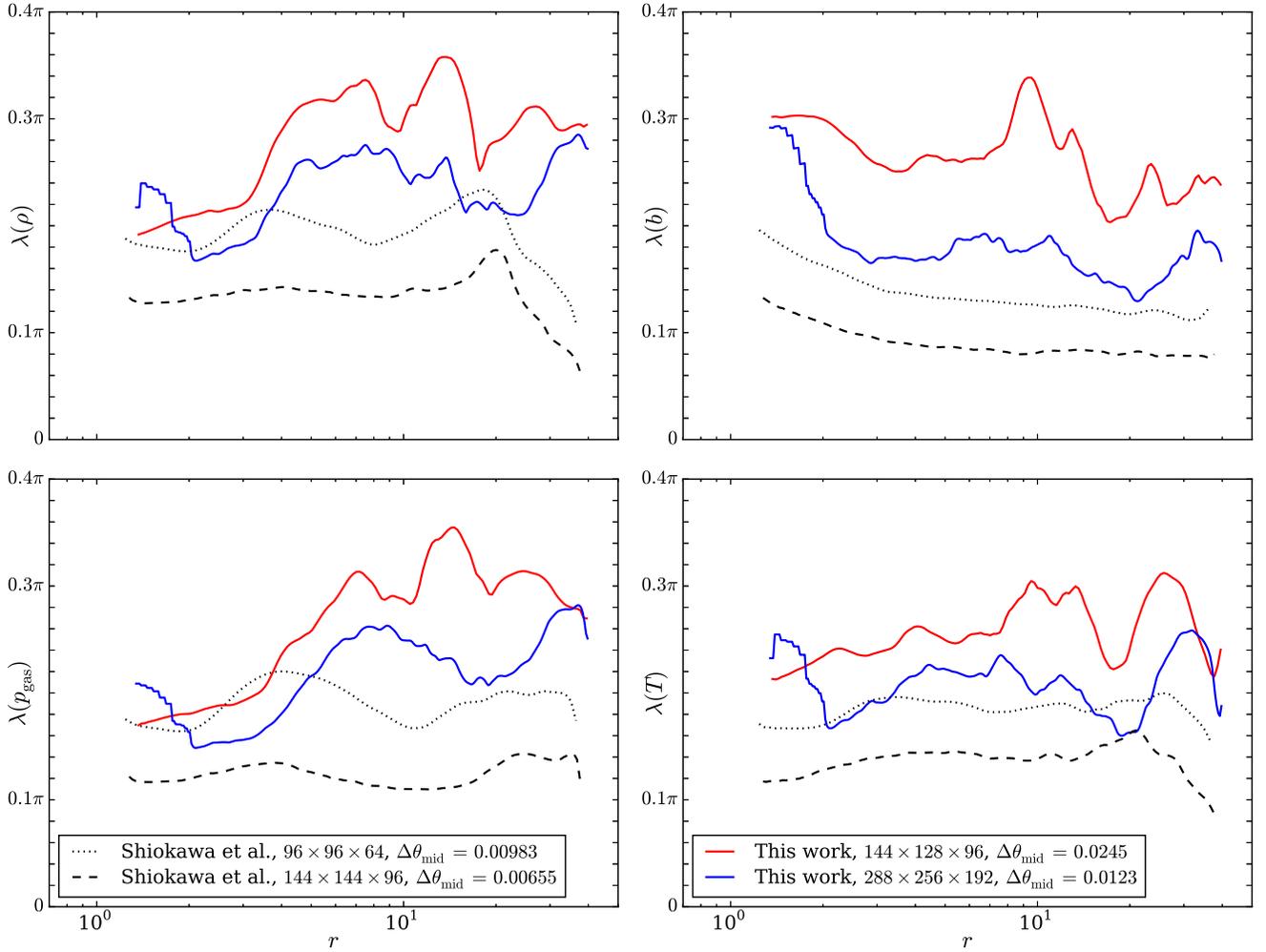}
  \caption{Azimuthal correlation lengths as functions of radius in the 3D torus runs. These are for density (top left), magnetic field (top right), gas pressure (equivalently internal energy; bottom left), and electron temperature (bottom right). The black lines show two select datasets from \citet{Shiokawa2012}, both using the ``soft'' polar boundary condition. \label{fig:torus_correlation}}
\end{figure}

We plot the runs of four correlation lengths in Figure~\ref{fig:torus_correlation}, to be compared to Figure~4 of \citet{Shiokawa2012}, from which select data has been replotted here. Compared to \citeauthor{Shiokawa2012}\ our disks have larger correlation lengths, especially in the outer regions. This is a sign that we have not yet reached the fully saturated, turbulent state in that region. It is worth repeating, however, that our midplane resolution is not as high as in \citeauthor{Shiokawa2012}, and moreover our correlation lengths decrease as we increase resolution.

\section{Performance}
\label{sec:performance}

In writing Athena++, we paid close attention to the performance of the code. The following techniques proved useful in improving performance:
\begin{itemize}
  \item Minimizing expensive evaluations and memory access. We precompute trigonometric and other such expensive functions related to the stationary metric as much as possible. At the same time, we minimize memory usage by storing only 1D and 2D arrays of precomputed values, as discussed in \S\ref{sec:algorithm:storage}.
  \item Sweeping through arrays as they are laid out in memory. For example, the hydrodynamical primitive variables are stored in a $5 \times N_3 \times N_2 \times N_1$ array. All nested loops range over $N_1$ in the innermost loop, avoiding strides greater than unity and thus minimizing cache misses. This includes calculating the fluxes, where even the $x^2$- and $x^3$-fluxes are evaluated in the $x^1$-direction order.
  \item Vectorizing all innermost loops. While modern compilers can often automatically vectorize loops, even when that entails inlining functions defined in other compilation units, there are several failure modes we sought to address. For one, complicated functions sometimes require manual inlining (e.g.\ in the HLLD solver, the function to evaluate the residual given a guess for the contact pressure). Additionally, vectorization cannot happen if the loop we seek to vectorize has indeterminate loops within it, as happens with algorithms that iterate until convergence. In such cases we examine how many iterations are typically used in cases were convergence happens at all, enabling us to fix a constant number of iterations.
\end{itemize}

We measure hydrodynamics and MHD performance as a function of both the metric and the Riemann solver. For special relativity, we run a shock tube in Minkowski coordinates. We then run the same shock tube using the full GR framework but still in Minkowski coordinates. Finally, we run a Fishbone-Moncrief torus (\S\ref{sec:torus}) in Kerr-Schild coordinates with nonzero spin. All tests are 3D. For the Riemann solvers we use a non-frame-transforming LLF solver (LLF-NT, as used in many GRMHD codes), a frame-transforming LLF solver (LLF-T), and either HLLC or HLLD (both frame-transforming) as appropriate. The distinction between using or not using a frame transformation (\S\ref{sec:algorithm:transformation}) only applies to GR.

Tests were performed on a single core of an Intel Xeon E5-2670 ($2.6\ \mathrm{GHz}$ Sandy Bridge) processor. We report the number of cells updated per wall time second in Table~\ref{tab:speeds}.

\begin{deluxetable}{cccccccc}
  \tablecaption{Athena++ single-core performance ($10^5$ cell updates per second) \label{tab:speeds}}
  \tablewidth{0pt}
  \tablehead{\colhead{Riemann Solver} & \colhead{SR} & \colhead{GR:\ Minkowski} & \colhead{GR:\ Kerr-Schild}}
  \startdata
    \sidehead{Hydro}
      LLF-NT & $12.8\phn$ & $10.4\phn$ & $\phn3.82$ \\
      LLF-T  & \nodata    & $10.1\phn$ & $\phn4.28$ \\
      HLLC   & $\phn8.57$ & $\phn7.36$ & $\phn3.28$ \\
    \sidehead{MHD}
      LLF-NT & $\phn3.24$ & $\phn4.70$ & $\phn2.94$ \\
      LLF-T  & \nodata    & $\phn2.60$ & $\phn1.97$ \\
      HLLD   & $\phn1.23$ & $\phn1.13$ & $\phn1.22$
  \enddata
\end{deluxetable}

The decrease in performance from SR to GR with the Minkowski metric reflects the cost of using the GR variable inversion and wavespeed formulas. The full cost of GR in realistic but reasonable metrics is represented by the final column of numbers, where nontrivial Kerr-Schild geometric factors enter into most calculations.

For a fixed geometry and Riemann solver, MHD problems run at $1/4$ to $3/4$ the speed of pure hydrodynamics problems in both SR and GR. We also performed tests using HLLE solvers. These ran at $93\pm6\%$ the speed of the corresponding LLF tests, with similar accuracies.

\begin{figure}
  \centering
  \includegraphics[width=6in]{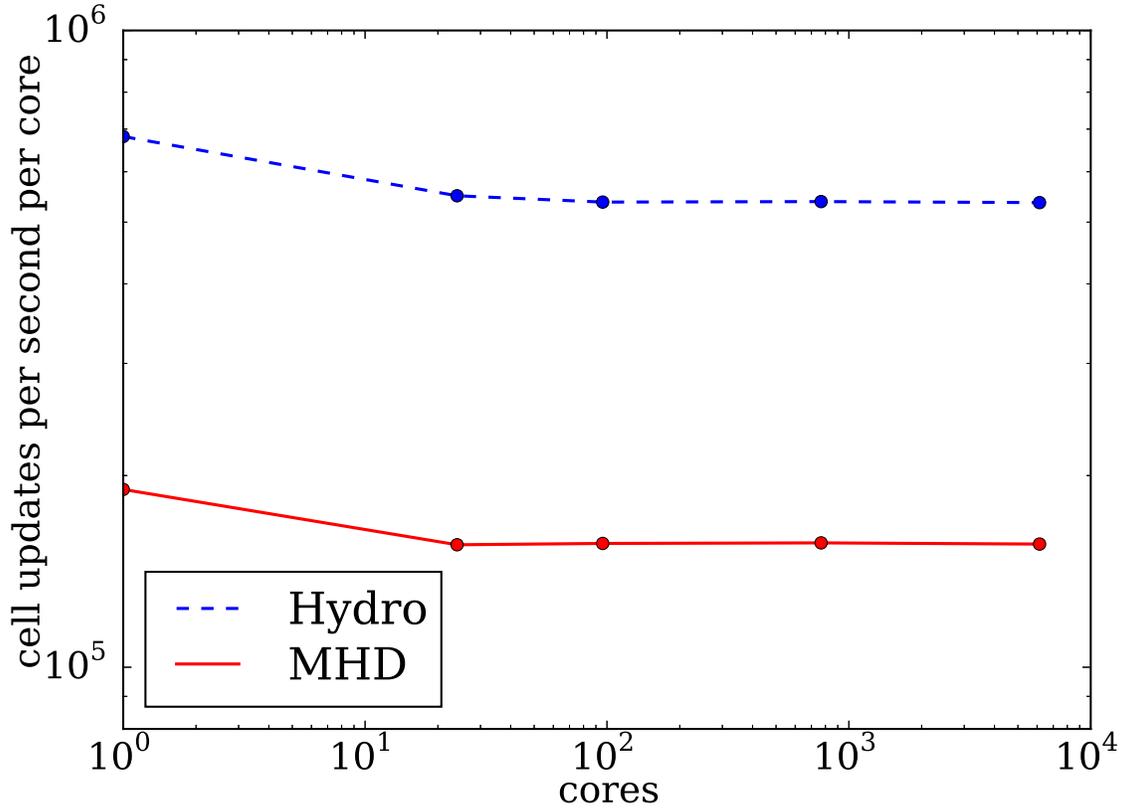}
  \caption{Performance per core running a 3D GRMHD simulation on a cluster. \label{fig:scaling}}
\end{figure}

We further report on the scaling of Athena++ with a full 3D GR computation using Kerr-Schild coordinates. Scaling was calculated on the NAOJ Cray~XC30, which has $24$ cores per node. We divide the domain into blocks of $64^3$ cells, assigning one block to each core. The scaling results out to $6144$ cores are shown in Figure~\ref{fig:scaling}. For both hydrodynamics and MHD, there is a $20\%$ per-core performance penalty to go from one core to one full node, almost certainly associated with saturating the memory bus when all cores on a node are used. However, the cost of going to many nodes from one is negligible. The performance of $6144$ cores is over $97\%$ relative to $24$ cores for hydrodynamics, and it is indistinguishable from $100\%$ for MHD.

\section{Summary}
\label{sec:summary}

We have described the GRMHD capabilities of the new Athena++ code, a modular framework for astrophysical fluid dynamics that now supports arbitrary stationary spacetimes. This code is distinguished by employing two techniques in particular that have been used with great success in Newtonian and special-relativistic settings but have not been combined in GRMHD. We support the use of advanced HLL Riemann solvers such as HLLC and HLLD (\S\ref{sec:algorithm:riemann}), employing a local frame transformation (\S\ref{sec:algorithm:transformation}) to achieve this without creating new general-relativistic Riemann solvers. Additionally, we support staggered-mesh CT in the style of \citet{Evans1988} and \citet{Stone2008} (\S\ref{sec:algorithm:ct}).

The modularity of the framework allows us to employ variations on individual parts of the implementation. In particular, we have static mesh refinement available, with adaptive mesh refinement being developed. We are also working to incorporate higher-order reconstruction such as the piecewise parabolic method of \citet{Colella1984}, as well as higher-order temporal integration such as the RK3 method of \citet{Shu1988}. Additionally, GR-compatible radiative transfer techniques are being developed for inclusion in the framework.

The code has been thoroughly tested. In addition to the linear waves, shock tubes, and blast waves used to test MHD codes in general, we employ tests specific to general relativity. These include reparameterizing Minkowski space with coordinates that induce time-space (\S\ref{sec:tests:linear}) or space-space (\S\ref{sec:tests:blast}) terms in the metric, as well as standard Schwarzschild Bondi accretion. We have further demonstrated that the code can properly simulate fluids around spinning black holes. This is true whether it is maintaining the stationary equilibrium of a hydrodynamical torus or evolving a turbulent, magnetized accretion flow. This latter problem combines hydrodynamics, magnetic fields, spatially varying time-space and space-space metric terms, nontrivial frame transformations with shifts orthogonal to and along interfaces, a number of geometrical source terms, and horizon-penetrating coordinates, thus exercising all components of the code in unison.

Finally, we note that the performance and scalability of GRMHD in Athena++ will enable us to perform numerical analyses not only with more accurate methods, but also with improved resolution. Numerous applications stand to benefit from this tool. For example, we look forward to better resolving magnetically arrested disks \citep{Tchekhovskoy2011} as well as disks tilted with respect to the black hole spin \citep{Fragile2007}.

\acknowledgments

We are particularly grateful to Vasileios Paschalidis for useful feedback, to Kengo Tomida for helping measure code performance, and to Hotaka Shiokawa for providing data used to compare our results to his. We acknowledge support from the NSF through grant AST-1333091. Many of the computations presented were performed with the resources made available via the Princeton Institute for Computational Science and Engineering.

\appendix

\section{Frame Transformation}
\label{sec:transformation}

Given a coordinate basis $\{\vec{\partial}_{(\mu)}\}$, we desire a new basis $\{\vec{e}_{(\hat{\mu})}\}$ satisfying the enumerated list of properties in \S\ref{sec:algorithm:transformation}, which we repeat here:
\begin{enumerate}
  \item \label{it:basis:orthogonal} $\vec{e}_{(\hat{\mu})}$ must be orthogonal to $\vec{e}_{(\hat{\nu})}$ for all $\mu \neq \nu$.
  \item \label{it:basis:normal} Each $\vec{e}_{(\hat{\mu})}$ must be normalized to have an inner product of $\pm1$ with itself, with $\vec{e}_{(\hat{t})}$ being timelike and $\vec{e}_{(\hat{\jmath})}$ being spacelike.
  \item \label{it:basis:time} $\vec{e}_{(\hat{t})}$ must be orthogonal to surfaces of constant $x^0$.
  \item \label{it:basis:projection} The projection of $\vec{e}_{(\hat{x})}$ onto the surface of constant $x^0$ is orthogonal to the surface of constant $x^i$ within that submanifold.
\end{enumerate}
As in the main text, we note that tensor components are indicated with unparenthesized subscripts and superscripts, with parentheses reserved for indices into generic sets. For concreteness we will work with an interface of constant $x^1$; results in the other directions follow from cyclic permutations $1 \mapsto 2 \mapsto 3$.

Property~\ref{it:basis:time} tells us
\begin{equation} \label{eq:e_t_low}
  e^{(\hat{t})}_\mu = A (1, 0, 0, 0)
\end{equation}
for some $A$, which is determined by property~\ref{it:basis:normal}:
\begin{equation}
  -1 = g^{\mu\nu} e^{(\hat{t})}_\mu e^{(\hat{t})}_\nu = A^2 g^{00}.
\end{equation}
Thus we have
\begin{equation}
  A = -(-g^{00})^{-1/2},
\end{equation}
where the sign is chosen such that the $\hat{t}$-direction is forward in time, as can be easily checked in the case $g_{\mu\nu} = \eta_{\mu\nu}$. Our first basis vector has components
\begin{equation} \label{eq:e_t}
  e_{(\hat{t})}^\mu = g^{\mu\nu} e^{(\hat{t})}_\nu = A g^{0\mu},
\end{equation}
and this is nothing more than the unit timelike normal $\hat{n}$ used in $3{+}1$ foliations of spacetime.

Next we set the $\hat{x}$ direction. It might be tempting to simply have it be parallel to $\vec{\pp}_{(1)}$, since this vector is already orthogonal to $\vec{e}_{(\hat{t})}$ by construction of the latter. However, this complicates the transformation of fluxes back to the global frame, as it results in $\hat{y}$- and $\hat{z}$-fluxes entering into the formula for $x^1$-fluxes. It is a more serious error to make $\vec{e}_{(\hat{x})}$ orthogonal to the $3$-surface of constant $x^1$. Though this would lead to a very simple formula for the flux conversion (requiring no knowledge of conserved quantities inside the wavefan), this vector can only be orthogonal to $\vec{e}_{(\hat{t})}$ if $g^{01}$ vanishes.

Instead, we enforce property~\ref{it:basis:projection}. This amounts to saying the projection of $\vec{e}_{(\hat{x})}$ onto the constant-$x^0$ subspace must have components
\begin{equation} \label{eq:e_x_3}
  \tensor*[^{(3)}]{e}{^{(\hat{x})}_i} \propto (1, 0, 0).
\end{equation}
Raising indices with the appropriate $3$-metric components $\gamma^{ij} = g^{ij} - g^{0i}g^{0j}/g^{00}$ and noting $e_{(\hat{x})}^0 = 0$ (by property~\ref{it:basis:orthogonal} and \eqref{eq:e_t_low}), we have
\begin{equation} \label{eq:e_x}
  e_{(\hat{x})}^\mu = B (g^{01}g^{0\mu} - g^{00}g^{1\mu}).
\end{equation}
The normalization is again set by property~\ref{it:basis:normal}:
\begin{equation}
  1 = g_{\mu\nu} e_{(\hat{x})}^\mu e_{(\hat{x})}^\nu = B^2 g^{00} (g^{00}g^{11} - g^{01}g^{01}).
\end{equation}
Choosing the sign such that $e_{(\hat{x})}^\mu = (0, +1, 0, 0)$ if $g_{\mu\nu} = \eta_{\mu\nu}$, we take
\begin{equation} \label{eq:factor_b}
  B = \big(g^{00} (g^{00}g^{11} - g^{01}g^{01})\big)^{-1/2}.
\end{equation}

Property~\ref{it:basis:orthogonal}, together with \eqref{eq:e_t_low}, forces
\begin{equation}
  e_{(\hat{y})}^0 = e_{(\hat{z})}^0 = 0;
\end{equation}
that is, $\vec{e}_{(\hat{y})}$ and $\vec{e}_{(\hat{z})}$ must lie in the constant-$x^0$ plane. From \eqref{eq:e_x_3} and property~\ref{it:basis:orthogonal} we have
\begin{equation}
  e_{(\hat{y})}^1 = e_{(\hat{z})}^1 = 0,
\end{equation}
which conveniently implies no $\hat{y}$- or $\hat{z}$-fluxes will be needed from the special relativistic Riemann solver.

For slight convenience we choose $e_{(\hat{z})}^2$ to be $0$. This corresponds to choosing an orientation of $\vec{e}_{(\hat{y})}$ and $\vec{e}_{(\hat{z})}$ in the plane they must span. Then we have
\begin{equation} \label{eq:e_z}
  e_{(\hat{z})}^\mu = C (0, 0, 0, 1).
\end{equation}
From property~\ref{it:basis:normal} we know
\begin{equation}
  1 = g_{\mu\nu} e_{(\hat{z})}^\mu e_{(\hat{z})}^\nu = C^2 g_{33},
\end{equation}
so
\begin{equation}
  C = (g_{33})^{-1/2},
\end{equation}
where again we check that we have a sensible definition when $g_{\mu\nu} = \eta_{\mu\nu}$: $e_{(\hat{z})}^\mu = (0, 0, 0, +1)$.

Finally, we solve for the two nonzero components of $\vec{e}_{(\hat{y})}$. Orthogonality (property~\ref{it:basis:orthogonal}) with $\vec{e}_{(\hat{z})}$, together with normalization (property~\ref{it:basis:normal}), constrain the solution to be
\begin{equation} \label{eq:e_y}
  e_{(\hat{y})}^\mu = D (0, 0, g_{33}, -g_{23}),
\end{equation}
where
\begin{equation}
  D = \big(g_{33} (g_{22}g_{33} - g_{23}g_{23})\big)^{-1/2}.
\end{equation}
Once again we check the sign choice: $e_{(\hat{y})}^\mu = (0, 0, +1, 0)$ if $g_{\mu\nu} = \eta_{\mu\nu}$.

Given the formulas \eqref{eq:e_t}, \eqref{eq:e_x}, \eqref{eq:e_z}, \eqref{eq:e_y}, it is a simple matter to assemble the transformation matrix \eqref{eq:to_global}:
\begin{equation}
  \tensor{M}{^\mu_{\hat{\nu}}} = e_{(\hat{\nu})}^\mu =
  \begin{pmatrix}
    A g^{00} & 0                               & 0                   & 0 \\
    A g^{01} & B (g^{01}g^{01} - g^{00}g^{11}) & 0                   & 0 \\
    A g^{02} & B (g^{01}g^{02} - g^{00}g^{12}) & \phantom{-}D g_{33} & 0 \\
    A g^{03} & B (g^{01}g^{03} - g^{00}g^{13}) & -D g_{23}           & C
  \end{pmatrix},
\end{equation}
with
\begin{subequations} \begin{align}
  A & = -(-g^{00})^{-1/2}, \\
  B & = \big(g^{00} (g^{00}g^{11} - g^{01}g^{01})\big)^{-1/2}, \\
  C & = (g_{33})^{-1/2}, \\
  D & = \big(g_{33} (g_{22}g_{33} - g_{23}g_{23})\big)^{-1/2}.
\end{align} \end{subequations}
The inverse transformation $\eqref{eq:to_local}$ is the inverse matrix
\begin{equation}
  \tensor{M}{^{\hat{\mu}}_\nu} =
  \begin{pmatrix}
    -A                                     & 0                                      & 0                     & 0   \\
    B g^{01}                               & -B g^{00}                              & 0                     & 0   \\
    B^2E g^{00} / D g_{33}                 & B^2F g^{00} / D g_{33}                 & 1 / D g_{33}          & 0   \\
    (B^2/C) g^{00} (G + E g_{23} / g_{33}) & (B^2/C) g^{00} (H + F g_{23} / g_{33}) & (1/C) g_{23} / g_{33} & 1/C
  \end{pmatrix},
\end{equation}
with
\begin{subequations} \begin{align}
  E & = g^{01} g^{12} - g^{11} g^{02}, \\
  F & = g^{01} g^{02} - g^{00} g^{12}, \\
  G & = g^{01} g^{13} - g^{11} g^{03}, \\
  H & = g^{01} g^{03} - g^{00} g^{13}.
\end{align} \end{subequations}
Both matrices are lower-diagonal, mostly as a result of our enumerated properties but also thanks to our choice of rotation $e_{(\hat{z})}^2 = 0$.

\bibliographystyle{aasjournal}
\bibliography{references}

\begin{thebibliography}{}
\expandafter\ifx\csname natexlab\endcsname\relax\def\natexlab#1{#1}\fi

\bibitem[{Abramowicz \& Fragile(2013)}]{Abramowicz2013}
Abramowicz, M.~A., \& Fragile, P.~C. 2013, LRR, 16, doi:10.12942/lrr-2013-1

\bibitem[{Anderson {et~al.}(2006)Anderson, Hirschmann, Liebling, \&
  Neilsen}]{Anderson2006}
Anderson, M., Hirschmann, E.~W., Liebling, S.~L., \& Neilsen, D. 2006, CQGra,
  23, 6503

\bibitem[{Ant{\'o}n {et~al.}(2010)Ant{\'o}n, Miralles, Mart{\'i},
  Ib{\'a}{\~n}ez, Aloy, \& Mimica}]{Anton2010}
Ant{\'o}n, L., Miralles, J.~A., Mart{\'i}, J.~M., {et~al.} 2010, ApJS, 188, 1

\bibitem[{Ant{\'o}n {et~al.}(2006)Ant{\'o}n, Zanotti, Miralles, Mart{\'i},
  Ib{\'a}{\~n}ez, Font, \& Pons}]{Anton2006}
Ant{\'o}n, L., Zanotti, O., Miralles, J.~A., {et~al.} 2006, ApJ, 637, 296

\bibitem[{Beckwith \& Stone(2011)}]{Beckwith2011}
Beckwith, K., \& Stone, J.~M. 2011, ApJS, 193, 6

\bibitem[{Brackbill \& Barnes(1980)}]{Brackbill1980}
Brackbill, J., \& Barnes, D. 1980, JCoPh, 35, 426

\bibitem[{Colella \& Woodward(1984)}]{Colella1984}
Colella, P., \& Woodward, P.~R. 1984, JCoPh, 54, 174

\bibitem[{Dedner {et~al.}(2002)Dedner, Kemm, Kr{\"o}ner, Munz, Schnitzer, \&
  Wesenberg}]{Dedner2002}
Dedner, A., Kemm, F., Kr{\"o}ner, D., {et~al.} 2002, JCoPh, 175, 645

\bibitem[{{Del~Zanna} {et~al.}(2007){Del~Zanna}, Zanotti, Bucciantini, \&
  Londrillo}]{DelZanna2007}
{Del~Zanna}, L., Zanotti, O., Bucciantini, N., \& Londrillo, P. 2007, A\&A,
  473, 11

\bibitem[{Einfeldt(1988)}]{Einfeldt1988}
Einfeldt, B. 1988, SJNA, 25, 294

\bibitem[{Etienne {et~al.}(2015)Etienne, Paschalidis, Haas, M{\"o}sta, \&
  Shapiro}]{Etienne2015}
Etienne, Z.~B., Paschalidis, V., Haas, R., M{\"o}sta, P., \& Shapiro, S.~L.
  2015, CQGra, 32, 175009

\bibitem[{Etienne {et~al.}(2012)Etienne, Paschalidis, Liu, \&
  Shapiro}]{Etienne2012}
Etienne, Z.~B., Paschalidis, V., Liu, Y.~T., \& Shapiro, S.~L. 2012, PhRvD, 85,
  024013

\bibitem[{Evans \& Hawley(1988)}]{Evans1988}
Evans, C.~R., \& Hawley, J.~F. 1988, ApJ, 322, 659

\bibitem[{Faber \& Rasio(2012)}]{Faber2012}
Faber, J.~A., \& Rasio, F.~A. 2012, LRR, 15, doi:10.12942/lrr-2012-8

\bibitem[{Falle \& Komissarov(1996)}]{Falle1996}
Falle, S., \& Komissarov, S. 1996, MNRAS, 278, 586

\bibitem[{Fishbone \& Moncrief(1976)}]{Fishbone1976}
Fishbone, L.~G., \& Moncrief, V. 1976, ApJ, 207, 962

\bibitem[{Fragile {et~al.}(2007)Fragile, Blaes, Anninos, \&
  Salmonson}]{Fragile2007}
Fragile, P.~C., Blaes, O.~M., Anninos, P., \& Salmonson, J.~D. 2007, ApJ, 668,
  417

\bibitem[{Gammie(1999)}]{Gammie1999}
Gammie, C.~F. 1999, ApJ, 522, L57

\bibitem[{Gammie {et~al.}(2003)Gammie, McKinney, \& T{\'o}th}]{Gammie2003}
Gammie, C.~F., McKinney, J.~C., \& T{\'o}th, G. 2003, ApJ, 589, 444

\bibitem[{Gardiner \& Stone(2005)}]{Gardiner2005}
Gardiner, T.~A., \& Stone, J.~M. 2005, JCoPh, 205, 509

\bibitem[{Gardiner \& Stone(2008)}]{Gardiner2008}
---. 2008, JCoPh, 227, 4123

\bibitem[{Giacomazzo \& Rezzolla(2007)}]{Giacomazzo2007}
Giacomazzo, B., \& Rezzolla, L. 2007, CQGra, 24, S235

\bibitem[{Harten {et~al.}(1983)Harten, Lax, \& {van~Leer}}]{Harten1983}
Harten, A., Lax, P.~D., \& {van~Leer}, B. 1983, SIAMR, 25, 35

\bibitem[{Hawley {et~al.}(1984)Hawley, Smarr, \& Wilson}]{Hawley1984}
Hawley, J.~F., Smarr, L.~L., \& Wilson, J.~R. 1984, ApJ, 277, 296

\bibitem[{Komissarov(1999)}]{Komissarov1999}
Komissarov, S. 1999, MNRAS, 303, 343

\bibitem[{Komissarov(2004)}]{Komissarov2004}
---. 2004, MNRAS, 350, 1431

\bibitem[{Marder(1987)}]{Marder1987}
Marder, B. 1987, JCoPh, 68, 48

\bibitem[{Mignone(2014)}]{Mignone2014}
Mignone, A. 2014, JCoPh, 270, 784

\bibitem[{Mignone \& Bodo(2005)}]{Mignone2005}
Mignone, A., \& Bodo, G. 2005, MNRAS, 364, 126

\bibitem[{Mignone \& Bodo(2006)}]{Mignone2006}
---. 2006, MNRAS, 368, 1040

\bibitem[{Mignone \& McKinney(2007)}]{Mignone2007}
Mignone, A., \& McKinney, J.~C. 2007, MNRAS, 378, 1118

\bibitem[{Mignone {et~al.}(2009)Mignone, Ugliano, \& Bodo}]{Mignone2009}
Mignone, A., Ugliano, M., \& Bodo, G. 2009, MNRAS, 393, 1141

\bibitem[{Misner {et~al.}(1973)Misner, Thorne, \& Wheeler}]{MTW}
Misner, C.~W., Thorne, K.~S., \& Wheeler, J.~A. 1973, {Gravitation} (New York:
  W.~H. Freeman and Company)

\bibitem[{Mohseni {et~al.}(2014)Mohseni, Mendoza, Succi, \&
  Herrmann}]{Mohseni2014}
Mohseni, F., Mendoza, M., Succi, S., \& Herrmann, H. 2014, PhRvD, 90, 125028

\bibitem[{M{\"o}sta {et~al.}(2014)M{\"o}sta, Mundim, Faber, Haas, Noble, Bode,
  L{\"o}ffler, Ott, Reisswig, \& Schnetter}]{Mosta2014}
M{\"o}sta, P., Mundim, B.~C., Faber, J.~A., {et~al.} 2014, CQGra, 31, 015005

\bibitem[{Muhlberger {et~al.}(2014)Muhlberger, Nouri, Duez, Foucart, Kidder,
  Ott, Scheel, Szil{\'a}gyi, \& Teukolsky}]{Muhlberger2014}
Muhlberger, C.~D., Nouri, F.~H., Duez, M.~D., {et~al.} 2014, PhRvD, 90, 104014

\bibitem[{Noble {et~al.}(2006)Noble, Gammie, McKinney, \&
  {Del~Zanna}}]{Noble2006}
Noble, S.~C., Gammie, C.~F., McKinney, J.~C., \& {Del~Zanna}, L. 2006, ApJ,
  641, 626

\bibitem[{Pons {et~al.}(1998)Pons, Font, Ib{\'a}{\~n}ez, Mart{\'i}, \&
  Miralles}]{Pons1998}
Pons, J.~A., Font, J.~A., Ib{\'a}{\~n}ez, J.~M., Mart{\'i}, J.~M., \& Miralles,
  J.~A. 1998, A\&A, 339, 638

\bibitem[{Ramshaw(1983)}]{Ramshaw1983}
Ramshaw, J.~D. 1983, JCoPh, 52, 592

\bibitem[{Roe(1981)}]{Roe1981}
Roe, P. 1981, JCoPh, 43, 357

\bibitem[{Schneider {et~al.}(1993)Schneider, Katscher, Rischke, Waldhauser, \&
  Maruhn}]{Schneider1993}
Schneider, V., Katscher, U., Rischke, D., Waldhauser, B., \& Maruhn, J. 1993,
  JCoPh, 105, 92

\bibitem[{Shibata \& Sekiguchi(2005)}]{Shibata2005}
Shibata, M., \& Sekiguchi, Y. 2005, PhRvD, 72, 044014

\bibitem[{Shiokawa {et~al.}(2012)Shiokawa, Dolence, Gammie, \&
  Noble}]{Shiokawa2012}
Shiokawa, H., Dolence, J.~C., Gammie, C.~F., \& Noble, S.~C. 2012, ApJ, 744,
  187

\bibitem[{Shu \& Osher(1988)}]{Shu1988}
Shu, C., \& Osher, S. 1988, JCoPh, 77, 439

\bibitem[{Stone \& Gardiner(2008)}]{Stone2008}
Stone, J.~M., \& Gardiner, T.~A. 2008, ApJS, 178, 137

\bibitem[{Tchekhovskoy {et~al.}(2011)Tchekhovskoy, Narayan, \&
  McKinney}]{Tchekhovskoy2011}
Tchekhovskoy, A., Narayan, R., \& McKinney, J.~C. 2011, MNRAS, 418, L79

\bibitem[{T{\'o}th(2000)}]{Toth2000}
T{\'o}th, G. 2000, JCoPh, 161, 605

\bibitem[{{van~Leer}(1979)}]{VanLeer1979}
{van~Leer}, B. 1979, JCoPh, 32, 101

\bibitem[{Woosley(1993)}]{Woosley1993}
Woosley, S. 1993, ApJ, 405, 273

\bibitem[{Zanotti \& Dumbser(2015)}]{Zanotti2015}
Zanotti, O., \& Dumbser, M. 2015, PhFl, 27, 074105

\end{thebibliography}

\end{document}